\documentclass[fleqn,useAMS,usenatbib,usegraphicx]{mn2e}
\usepackage{amssymb,amsmath}

\def\aap{A\&A}%
\def\aaps{A\&AS}%
\def\apj{ApJ}%
\def\apjs{ApJ}%
\def\apjl{ApJ}%
\def\mnras{MNRAS}%
\def\icarus{Icarus}%
\def\planss{Planet.~Space~Sci.}%

\newcommand{\se}[1]{Section~\ref{sec:#1}}

\newcommand{\Se}[1]{\mbox{Section\ \ref{sec:#1}}}
\newcommand{\eq}[1]{equation~(\ref{eq:#1})}
\newcommand{\ineq}[1]{inequality~(\ref{eq:#1})}
\newcommand{\identity}[1]{\mbox{identity~(\ref{eq:#1})}}
\newcommand{\solut}[1]{solution~(\ref{eq:#1})}
\newcommand{\eqp}[1]{equation~\ref{eq:#1}}
\newcommand{\Eq}[1]{Equation~(\ref{eq:#1})}
\newcommand{\fg}[1]{Fig.~\ref{fig:#1}}
\newcommand{\Eqs}[2]{Equations\ (\ref{eq:#1}), (\ref{eq:#2})}
\newcommand{\eqs}[2]{equations\ (\ref{eq:#1}), (\ref{eq:#2})}

\newcommand{\eqsto}[2]{equations\ (\ref{eq:#1})--(\ref{eq:#2})}

\newcommand{\Fg}[1]{Figure~\ref{fig:#1}}
\newcommand{\Tb}[1]{\mbox{Table\ \ref{tab:#1}}}
\newcommand{\app}[1]{\mbox{Appendix\ \ref{app:#1}}}
\newcommand\cf{cf.}
\newcommand\eg{e.g.,}
\newcommand\ie{i.e.,}

\newcommand\etc{etc}
\newcommand{\e}{\bmath{e}}
\newcommand{\Ei}{\mathrm{Ei}}
\newcommand{\Si}{\,\mathrm{Si}}
\newcommand{\Ci}{\,\mathrm{Ci}}
\newcommand\mEarth{\mathrm{M}_\oplus}

\usepackage{color}

\title{The steady-state flow pattern past gravitating bodies}
\author[C.W.~Ormel]{C.W.~Ormel$^{1}$\thanks{E-mail:
ormel@berkeley.edu}\thanks{Hubble Fellow}\\
$^{1}$Astronomy Department, University of California, Berkeley, CA 94720, USA}
\begin{document}
\maketitle
\label{firstpage}
\begin{abstract}
Gravitating bodies significantly alter the flow pattern (density and
velocity) of the gas that attempts to stream past. Still, small protoplanets in
the Mars--super-Earth range can only bind limited amounts of nebular gas; until the
so-called critical core mass has been reached ($\sim$1--10 Earth masses) this
gas is in near hydrostatic equilibrium with the nebula. Here we aim for a general
description of the flow pattern surrounding these low-mass, embedded
planets. Using various simplifying assumptions (subsonic, 2D, inviscid flow,
\etc), we reduce the problem to a partial differential equation that we solve
numerically as well as approximate analytically. It is found that the boundary
between the atmosphere and the nebula gas strongly depends on the value of the
disc headwind (deviation from Keplerian rotation). With increasing headwind the atmosphere 
decreases in size and also becomes more asymmetrical.
Using the derived flow pattern for the gas, trajectories 
of small solid particles, which experience both gas drag and gravitational forces,
are integrated numerically. Accretion rates for small
particles (dust) are found to be low, as they closely follow the streamlines, which
curl away from the planet. However, pebble-size particles achieve large
accretion rates, in agreement with previous numerical and analytical works.
\end{abstract}


\section{Introduction}
\label{sec:intro}
Accretion of gas by gravitating objects is a common theme in astrophysics. The pioneering work in this field goes back the works of \citet{HoyleLyttleton1939} and \citet{BondiHoyle1944}, who analytically derived the accretion rates for point-bodies (together known as Bondi-Hoyle-Lyttleton or BHL-accretion). 
Follow-up studies have refined these estimates, by considering the effects of the Mach number of the incident flow and the equation of state for the gas, mostly for supersonic flows \citep[\eg][]{Hunt1979,ShimaEtal1985,Ruffert1995}. In astrophysics, BHL-accretion plays a role in Binary systems, star- and galaxy formation, and the formation of gas-rich planets \citep[\eg][]{KleyEtal1995,BonnellEtal2001,NelsonBenz2003}. 

According to the core accretion model of planet formation, planets start out small and rocky as protoplanetary embryos \citep{Mizuno1980,PollackEtal1996,Weidenschilling1997,HubickyjEtal2005}. Once these embryos reach a size where their Bondi radius 
\begin{equation}
  \label{eq:R-bondi}
  R_b 
  \equiv \frac{G_N M_p}{c_\infty^2}
\end{equation}
starts to exceed their physical radius $R$, they start to bind the gas from the nebula, forming a putative atmosphere. 
In \eq{R-bondi} $G_N$ is Newton's constant, $M_p$ the mass of the perturber, and $c_\infty$ the sound speed of the unperturbed system. As accreting planets are hotter than the background nebula -- either from infalling solids (planetesimals) or from the residual heat of their formation \citep{IkomaHori2012} -- the resulting pressure gradient prevents contraction of the atmosphere. Therefore, the amount of gas these (small) bodies can acquire is initially limited: the protoplanet's atmosphere is in pressure-equilibrium with the surrounding nebular gas. \footnote{Indeed, in a numerical experiment \citet{TerquemHeinemann2011} recently showed that, starting from a much condensed configuration, a protoplanet's atmosphere will expands and reconnect with the disc's gas to restore the near-hydrostatic state.} Small, hot protoplanets do not acquire gas akin to the BHL-regime: they must cool first.

However, at some critical core mass $M_\mathrm{crit}$ the steady state picture is no longer appropriate. Many detailed, 1D models have been developed to describe the quasi-static density structure of the protoplanet atmosphere and the cross-over point \citep{Stevenson1982,Wuchterl1993,InabaIkoma2003,Rafikov2006}. This crossover mass is often quoted as $\sim$$10\ \mEarth$ (Earth masses) but its precise value depends on detailed atmosphere models, and thereby on the equation of state, accretion luminosity, and opacities of gas and grains; it decreases to $\sim$1~$\mEarth$ for grain-free atmospheres \citep{HoriIkoma2010}. Obviously, understanding the processes that determine $M_\mathrm{crit}$ is of key importance for planet formation; it may determine the transition between (super)-Earth/Neptune-like planets (bodies that have acquired limited amounts of gas) and gas giants (bodies that accrete gas in the BHL-regime).

In this work we will focus on the small planet regime, assuming a quasi-steady state. In a disc, the unperturbed gas flow, as seen from a frame rotating with the planet, consist of two components: a \textit{shear} (due to the [Keplerian] rotation of the disc) and a systematic offset, which we refer to as the \textit{headwind}. The latter arises from the slightly subkeplerian rotation of the gas, which is partially support by pressure. For a planet on a circular orbit, the value of the headwind is $\eta v_k$ with $\eta$ \citep{AdachiEtal1976,Weidenschilling1977}:
\begin{equation}
  \eta 
  = \frac{1}{2 \rho a \Omega^2} \frac{\partial P}{\partial a} 
  = -\frac{1}{2} \frac{c_s^2}{v_K^2} \frac{\partial \ln P}{\partial \ln a}
  \sim \left( \frac{c_s}{v_k} \right)^2
  \label{eq:vhw}
\end{equation}
where $a$, $\rho$, $P$, $\Omega$, and $v_K$ are the disc radius (semi-major axis), the corresponding gas density, pressure, orbital frequency, and orbital velocity. In this work the headwind is also defined in terms of a Mach number, $v_\mathrm{hw}={\cal M}_\mathrm{hw}c_\infty$, \ie\ ${\cal M}_\mathrm{hw}=\eta v_k/c_s \simeq \eta^{1/2}$. Only positive values of $v_\mathrm{hw}$ (a headwind; the gas rotates lower than Keplerian) are consider, but we remark that in special locations (`pressure bumps') $v_\mathrm{hw}$ may reverse sign. In addition, the flow, as seen in the frame of the planet, is subject to (noninertial) Coriolis forces and to the tidal force from the distant star. The flow pattern is therefore quite rich. 


Understanding the flow pattern close to the planet is important for several reasons. For example, the flow pattern in the co-orbital region of the planet, where streamlines make a `U-turn' (the horseshoe) is a critical ingredient for the co-orbital torque (\citealt{Ward1991}; see also \citealt{PaardekooperPapaloizou2009}). Also, the location of the boundary between the nebular gas and the protoplanet's atmosphere is a parameter that affects the thermal evolution of (growing) protoplanets \citep{LissauerEtal2009}.  The amount of gas that can be bound during the protoplanetary disc phase affects the long-term evolution of super-Earths and mini-Neptunes \citep{IkomaGenda2006,IkomaHori2012,LopezEtal2012}.

The gravitationally perturbed flow pattern also affects the behavior of particles. \citet{WeidenschillingDavis1985} and \citet{Paardekooper2007} investigated the accretion potential of (small) particles.  The secular particle-planet interaction has been studied more generally by \citet{MutoInutsuka2009}. Of special importance in the context of this work is the new `pebble-accretion' mechanism, where protoplanets quickly accrete small particles, as found by both numerical simulations \citep{LambrechtsJohansen2012,MorbidelliNesvorny2012} as well as analytical estimates \citep{OrmelKlahr2010,PeretsMurray-Clay2011}. The latter studies however did not account for the modification of the gas flow by the gravity of the protoplanet -- the topic of this paper.

The goal of this paper is to obtain a quantitative description of the density structure and the flow (gas velocity) in the vicinity of a gravitating body -- in particularly, a protoplanet in a Keplerian-rotating disc --  and to assess the role of key parameters like: the headwind, shear, and planet mass.  The low-mass planet regime under consideration is characterized by the following scale hierarchy:
\begin{equation}
  \label{eq:hierarchy}
  R < R_b < R_h < H,
\end{equation}
where $H$ is the scaleheight of the disc, $R$ the radius of perturber, and $R_h$ the Hill radius
\begin{equation}
  \label{eq:R-Hill}
  R_h 
  \equiv a_0 \left( \frac{M_p}{3M_\star} \right)^{1/3}
\end{equation}
(with $M_\star$ the stellar mass) -- the scale on which the solar gravity rivals that of the perturber.  At the lower range, $R\simeq R_b$, bodies are about $\sim 10^3$ km in size, or $10^{-3}$--$10^{-2}$ $\mEarth$ in mass. With increasing mass both $R_b$ and $R_h$ increase, but the ratio $R_h/R_b$ decreases. At the high-mass end of \ineq{hierarchy} $R_b\simeq R_h \simeq H$. This corresponds to masses of $\sim$10 $\mEarth$, somewhat larger for the outer disc.

One key approximation that is employed is to neglect contributions originating from the Lindblad region, \ie\ the distance $x\sim H$, where shock waves are excited. Resolving the shock at these distances is important for calculating the torque that is exerted on the planets, which determines the orbital evolution (migration) of the planet. However, as an approximation for the flow close to the planets, \ie\ on scales $\ll$$H$, \citet{PaardekooperPapaloizou2009} argued that it is justifiable to omit the contributions arising from the Lindblad torque region.  That is, they showed that this procedure, referred to as `torque cut-off', is permissible -- up to levels of $\sim$10\% accuracy -- provided that the softening radius, which is customary applied to smoothen the gravitational potential in numerical simulations, is chosen small enough. In this study no softening is present; instead the numerical setup is characterized by a \textit{surface} (inner boundary condition). Avoiding the softening is advantageous, since it is a parameter that, unless carefully chosen, could affect the outcome of the numerical experiment \citep{DongEtal2011,MuellerEtal2012}.

These considerations imply that the flow past the protoplanet is subsonic.  
A further assumption is that the flow is two dimensional (2D), which allows us to formulate the flow in terms of a single scalar quantity, the stream function $\Psi$. Employing the 2D assumption significantly reduces the complexity of the problem \citep[\eg][]{KorycanskyPapaloizou1996}. Extending the stream function formulation to 3D configurations is straightforward for axisymmetric flows \citep{LeeStahler2011}, but much more difficult -- often unpractical -- for truly 3D flows. Similar simplifying assumptions are that the flow is inviscid and steady (non-turbulent). These idealizations allow us to conduct a thorough parameter study for the flow pattern in the vicinity of the perturbing body, making a systematic investigating of the sensitivity of the flow pattern to the various parameters involved (protoplanet mass, nebular headwind, equation of state, numerical parameters) possible. 

The structure of the paper is as follows. In \se{streamf} we formulate the problem and the underlying (inviscid) fluid equations. In \se{linear} a first analytic model for the flow pattern is presented by a linear perturbation analysis.  \Se{simul-res} highlights the key results obtained from a numerical parameter study.  \Se{approx} a more complete analytic model is given, which describes the critical atmosphere region, under the approximation that the density is a radial function. In \se{p-traj} the analytic model for the gas flow is used to numerically integrate trajectories of small particles in order to assess their accretion behavior.  Further implications are discussed in \se{discussion}. \Se{conclusions} presents the conclusions.

\section{Stream function formulation}
\label{sec:streamf}
In this section the equations of continuity and force-balance (Euler) are rewritten in terms of Bernoulli's equation and a diffusion equation for the stream function $\Psi$ -- the quantity that together with the surface density $\Sigma$ characterizes the flow. The numerical solution to these equations is presented in \se{simul-res} and an analytical approximation is given in \se{approx}. 
A key feature of the approach is to connect the local solution near the perturber to the unperturbed solutions at large radii -- the far field. The far field flow pattern is assumed to be known and specified by the following constant quantities: surface density ($\Sigma_\infty$), sound speed ($c_\infty$), vorticity ($\omega_\infty$), and headwind ($v_\mathrm{hw}$).

\subsection{The stream function and far field solution}
The 2D shearing-sheet approximation is adopted, with the centre of the coordinate frame rotating at an angular frequency equal to the local orbital frequency $\Omega_0$. In the absence of a perturbing body the flow is, without loss of generality, assumed to be directed in the negative $y$-direction:
\begin{equation}
  \bmath{v}_\infty = (-v_\mathrm{hw} -q\Omega_0 x) \e_y = (-v_\mathrm{hw} +w_\infty x) \e_y,
  \label{eq:vinf}
\end{equation}
where $q$ is the dependence of $\Omega$ on disc radius, $\Omega \propto a^{-q}$ at $a=a_0$. All subscripts involving `$\infty$' refer to the far field (unperturbed) solution. Instead of $q$ the linear shear can be quantified by the vorticity in the far field:
\begin{equation}
  \bmath{\omega}_\infty 
  = \omega_\infty \bmath{e}_z = \nabla \bmath{\times} \bmath{v}_\infty = -q\Omega_0\bmath{e}_z.
  \label{eq:winf}
\end{equation}
In a Keplerian disc, $q=3/2$ and $w_\infty=-3\Omega_0/2$. 

The 2D stream function, $\bmath{\Psi}=\Psi \bmath{e}_z$, is defined such that it satisfies the continuity equation in steady state, $\nabla \cdot \Sigma \bmath{v} = 0$:
\begin{equation}
  \Sigma \bmath{v} \equiv \nabla \bmath{\times} \Psi\bmath{e}_z.
  \label{eq:continuity}
\end{equation}
In the far field, $\Sigma = \Sigma_\infty$ is constant and $\Psi$ becomes
\begin{equation}
  \label{eq:psiinf}
  \Psi_\infty 
  = \left( v_\mathrm{hw} x -\frac{1}{2}\omega_\infty x^2 \right)\Sigma_\infty.
\end{equation}
Dividing the continuity \eq{continuity} by $\Sigma$ and taking the curl we obtain the vorticity of the flow:
\begin{equation}
  \label{eq:diffPsi}
  \omega = \bmath{e}_z \cdot \nabla \bmath{\times} \bmath{v}
  = \bmath{e}_z \cdot \left( \nabla \bmath{\times} \frac{ \nabla \bmath{\times} \Psi }{\Sigma} \right)
  = -\nabla \cdot (\frac{1}{\Sigma} \nabla \Psi),
\end{equation}
where we used the (general) vector identity $\bmath{e}_z \cdot \nabla \bmath{\times} \bmath{A} = -\nabla \cdot (\bmath{e}_z \bmath{\times} \bmath{A})$ and the 2D-specific
\begin{equation}
  \nabla \Psi = \bmath{e}_z \bmath{\times} (\nabla\bmath{\times} \Psi\bmath{e}_z).
  \label{eq:Psi-2D-id}
\end{equation}
Furthermore, the vortensity $(2\Omega_0+\omega)/\Sigma$, is conserved along streamlines.\footnote{See \eg\ \citet{LandauLifshitz1959} for a proof. Morover, conservation of vortensity implies that the flow is barotropic (\eqp{enthalpy}).} Since $\Sigma$ and $\omega$ are constant in the far field, the vortensity is constant everywhere and $\omega$ can be expressed in terms of $\Sigma$:
\begin{equation}
  \omega = (2\Omega_0 +\omega_\infty)\frac{\Sigma}{\Sigma_\infty} -2\Omega_0;
  \label{eq:vorticity}
\end{equation}
and inserted in \eq{diffPsi} to obtain
\begin{equation}
  \nabla \cdot \frac{1}{\Sigma} \nabla \Psi 
  = 
  -\tilde\omega \frac{\Sigma}{\Sigma_\infty} +2\Omega_0,
  \label{eq:Diffusion}
\end{equation}
where $\tilde\omega = 2\Omega_0 +\omega_\infty$ ($=+\Omega_0/2$ in a Keplerian disc).  Thus, $\Psi$ obeys a diffusion equation in which both the diffusion coefficient ($1/\Sigma$) and the source term (the RHS of \eqp{Diffusion}) are functions of $\Sigma$. If the background flow is shear-free (headwind only) the source term is zero. If in addition the flow is incompressible Laplace equation is obtained: $\nabla^2 \Psi = 0$. For the general case, however, both $\Psi$ and $\Sigma$ are unknown. 

\subsection{Bernoulli's equation}
\label{sec:Bernoulli}
To solve the degeneracy between $\Sigma$ and $\Psi$, \eq{Diffusion} is supplemented by the Euler equation (force balance).  In the uniformly \textit{rotating} reference system that is considered here, the Euler equation contains fiducial forces:
\begin{equation}
  \label{eq:Euler}
  \frac{\partial \bmath{v}}{\partial t} +\bmath{v} \cdot \nabla \bmath{v} 
  = -2\bmath{\Omega}_0\bmath{\times}\bmath{v} -\bmath{\Omega}_0\bmath{\times}(\bmath{\Omega}_0\bmath{\times}\bmath{r}) -\frac{1}{\Sigma}\nabla P - \nabla \Phi_g,
\end{equation}
where $\Phi_g$ is the combined gravitational potential (of the planet and the star) acting on the fluid element.  The pressure gradient term $\nabla P/\Sigma$ is split into a global contribution, which gives rise to the headwind, and a local perturbation caused by the planet. The global term equals $(dP_\infty/dr)/\Sigma_\infty \equiv -2\eta v_k \Omega_0 = -2v_\mathrm{hw}\Omega_0$ (see \eqp{vhw}) and is added to \eq{Euler} such that $\nabla P/\Sigma$ denotes the perturbation. In addition, it can be shown that the centrifugal force term ($-\bmath{\Omega}_0\bmath{\times}(\bmath{\Omega}_0\bmath{\times}\bmath{r}) = -\Omega_0 r^2$) and the solar force term ($\nabla \Phi_\mathrm{sun}$) combine into $2q\bmath{\Omega_0} \bmath{\times} \bmath{e}_x$.
Since by assumption the flow is steady ($\partial/\partial t = 0$),  \eq{Euler} reads
\begin{equation}
  \label{eq:Euler2}
  \bmath{v} \cdot \nabla \bmath{v} 
  = -2\bmath{\Omega}_0\bmath{\times}\bmath{v} -\frac{1}{\Sigma}\nabla P +(2q\Omega_0^2 x +2v_\mathrm{hw}\Omega_0)\bmath{e}_x -\nabla \Phi_P,
\end{equation}
where $\Phi_P$ is solely due to the gravitating body at the centre of the reference frame and $\nabla P$ corresponds to the pressure perturbations induced by it. In the far field the LHS as well as the pressure and potential terms on the RHS of \eq{Euler2} vanish; and it can be verified that \eq{vinf} is a solution to \eq{Euler2}.

Using the identity 
$\frac{1}{2} \nabla v^2 = \bmath{v}\cdot \nabla \bmath{v} + \bmath{v} \bmath{\times} \nabla \bmath{\times} \bmath{v}$
\eq{Euler2} transforms into:
\begin{align}
  \label{eq:Euler3}
  2\bmath{\Omega_0}\bmath{\times}\bmath{v} -\bmath{v} \bmath{\times} \nabla \bmath{\times} \bmath{v} 
  =& -\frac{1}{2} \nabla v^2 -\frac{1}{\Sigma}\nabla P \\
   & -\nabla \left( \Phi_P -q\Omega_0 x^2 -2v_\mathrm{hw}\Omega_0 x \right),
   \nonumber
\end{align}
or, in terms of the vorticity $\bmath{\omega}=\nabla \bmath{\times} \bmath{v}$:
\begin{equation}
  \label{eq:Bernoulli}
  \bmath{v} \bmath{\times} (\bmath{\omega} +2\Omega_0\bmath{e}_z) 
  \equiv \nabla B
\end{equation}
with $B$ the Bernoulli function
\begin{equation}
  B
  = \frac{1}{2}v^2 +W +\Phi_P +w_\infty\Omega_0x^2 -2v_\mathrm{hw}\Omega_0x.
  \label{eq:Bconstant}
\end{equation}
In \eq{Bernoulli} it was further assumed that the flow is barotropic:
\begin{equation}
  \frac{\nabla P}{\Sigma}
  = \nabla \int \frac{dP}{\Sigma}
  \equiv \nabla W
  \label{eq:enthalpy}
\end{equation}
with $W$ the enthalpy. In the far field $W$ is assumed to be zero, $W(\Sigma_\infty)=0$, and the Bernouilli function reduces to
\begin{align}
  \label{eq:Binf}
  B_\infty 
  =&\ \frac{1}{2}v_\infty^2 +w_\infty \Omega_0 x^2 -2v_\mathrm{hw}\Omega_0 x \\ \nonumber
  =&\ \frac{1}{2}v_\mathrm{hw}^2 -v_\mathrm{hw}(2\Omega_0+\omega_\infty)x +\frac{1}{2}\omega_\infty(2\Omega_0+\omega_\infty)x^2
\end{align}
where \eq{vinf} has been inserted for $v_\infty$.

Since the LHS of \eq{Bernoulli} is orthogonal to $\bmath{v}$, it vanishes upon integration along a streamline.  This implies that $B$ is conserved along streamlines (Bernoulli equation).  To find the relation between $B$ and $\Psi$, \eq{continuity} is inserted into \eq{Bernoulli}. The LHS of \eq{Bernoulli} is then written in terms of $\Psi$:
\begin{equation}
  \label{eq:PsiBlink}
  \frac{\nabla \bmath{\times} \Psi\bmath{e}_z}{\Sigma} \bmath{\times} \frac{\Sigma}{\Sigma_\infty}\tilde{w}\bmath{e}_z 
  = -\frac{\tilde{w}}{\Sigma_\infty} \nabla \Psi
  = \nabla B,
\end{equation}
where \eq{vorticity} and \identity{Psi-2D-id} are used. As $\tilde\omega$ and $\Sigma_\infty$ are constant, \eq{PsiBlink} simply implies that $B$ is a linear function of $\Psi$:
\begin{equation}
  B(\Psi) 
  = 
  -\frac{\tilde{w}}{\Sigma_\infty}\Psi + \frac{1}{2}v_\mathrm{hw}^2,
  \label{eq:Bnew}
\end{equation}
where the integration constant has been defined such that $B(\Psi_\infty) = B_\infty$ (\eqp{Binf}).

Together, \eqs{Bconstant}{Bnew} provide a second relation between $\Psi$ and $\Sigma$:
\begin{equation}
  \label{eq:Bern1}
  \frac{1}{2}\frac{|\nabla \Psi|^2}{\Sigma^2}
  -\frac{v_\mathrm{hw}^2}{2}
  +\frac{\tilde\omega}{\Sigma_\infty}\Psi
  +W
  +\Phi_P +w_\infty\Omega_0 x^2 -2v_\mathrm{hw}\Omega_0x = 0.
\end{equation}

\subsection{Units and nondimensionalisation}
\label{sec:units}
\begin{figure}
  \centering
  \includegraphics[width=84mm]{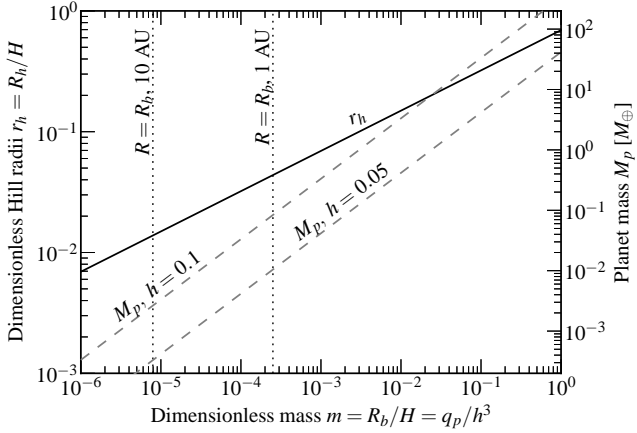}
  \caption{Left $y$-axis: the dimensionless Hill radius $r_h$ as function of mass parameter $m\equiv R_b/H$ (solid line). Right $y$-axis: the perturber mass $M_p$ corresponding to $m$ for two values of the disk aspect ratio $h=H/a_0$ (dashed lines). For the conversion of $q_p$ to $M_p$ the star was assumed to be of solar mass. Vertical dotted lines show the mass where $R=R_b$, the point where the embryo starts to bind the nebular gas, at distances of 1 AU and 10 AU. }
  \label{fig:dim-quants}
\end{figure}
For our problem it is natural to measure frequencies ($\omega,\tilde\omega$ and $\Omega_0$) in terms of $\Omega_0$ and densities in terms of $\Sigma_\infty$. A natural length unit is not as obvious, since both the Bondi radius, the Hill radius, and the scaleheight of the disk $H$ could qualify. Here we adopt the scaleheight as the length unit, which implies that the unit of velocity is $c_\infty$. The dimensionless disk headwind ${\cal M}_\mathrm{hw} = v_\mathrm{hw}/c_s$ is referred to as the Mach number of the headwind. Because $R_b$ scales with the perturber mass, the dimensionless Bondi unit is also a dimensionless mass:
\begin{equation}
  m 
  \equiv \frac{R_b}{H} 
  = \frac{GM_p}{H^3\Omega_0^2}
  = \frac{q_p}{ h^3}
  \label{eq:m-def}
\end{equation}
where $q_p=M_p/M_\star$ and $h=H/a_0$ the aspect ratio of the disk. We will use $m$ as the parameter for the mass of the perturbing body. The dimensionless Hill radius can be written
\begin{equation}
  r_h 
  = \frac{R_h}{H}
  = \left( \frac{q_p}{3h^3} \right)^{1/3}
  = (m/3)^{1/3}.
\end{equation}
\Fg{dim-quants} shows the relation among these quantities as function of the mass parameter $m$. The regime $m \ll 1$ applies in this paper (low mass limit). Small bodies start to bind an atmosphere when their physical radius starts to exceed the Bondi radius, $R\simeq R_b$. This correspond to an $m$-value of $m\sim \sqrt{M_\star/\rho_s a_0^3}$ where $\rho_s$ is the internal density of the body. With increasing mass the Bondi and Hill radius start to converge on each other. The point $m\sim1$ signifies the transition to the high mass regime (for which $R_b>R_h>H$) which is not considered here.



In nondimensional units the diffusion \eq{Diffusion} and Bernoulli \eq{Bern1} read:
\begin{equation}
  \nabla \cdot \frac{1}{\Sigma} \nabla \Psi = -\tilde\omega \Sigma +2
  \label{eq:D-nondim}
\end{equation}
\begin{equation}
  \frac{1}{2}\frac{|\nabla \Psi|^2}{\Sigma^2}
  +W(\Sigma)
  =
  -\tilde\omega\Psi
  +\frac{m}{r}
  +\frac{ {\cal M}_\mathrm{hw}^2}{2}
  -w_\infty x^2 +2{\cal M}_\mathrm{hw}x.
  \label{eq:B-nondim}
\end{equation}
These are two equations for the two unknown quantities, $\Sigma$ and $\Psi$.  The unperturbed solution ($\Sigma=\Sigma_\infty=1$; $\Psi=\Psi_\infty$) is obtained when $m=0$, as it was specified that $W(\Sigma=1)=0$.

\begin{figure}
  \centering
  \includegraphics[width=84mm]{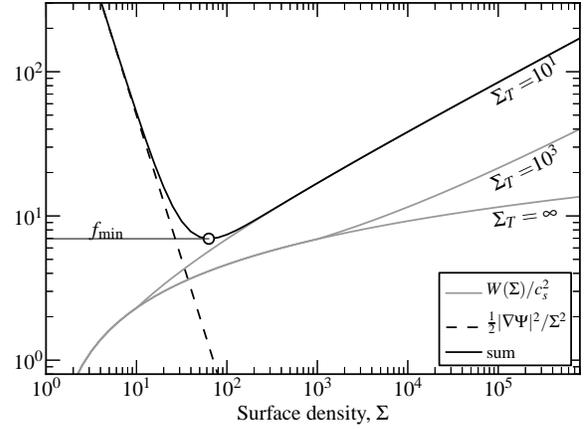}
  \caption{(gray curves) The enthalpy function $W$ (\eqp{W-expr}) as function of density for several values of the transition density: $\Sigma_T = 10, 10^3$ and $\infty$ (isothermal). (black curve) the LHS of \eq{B-nondim} for a value for $\frac{1}{2}|\nabla \Psi|^2=10^4$. The minimum density $f_\mathrm{min}$ for which the Bernoulli equation will obtain real solutions is indicated by an open circle.}
  \label{fig:wmodel}
\end{figure}
\subsection{Equation of state (EOS)}
These equations are supplemented by an equation of state (EOS). We adopt:
\begin{equation}
  \label{eq:eos}
  P = \left\{ \begin{array}{ll}
    \displaystyle
    c_\infty^2 \Sigma                                                & (\Sigma \le \Sigma_T) \\
    \displaystyle
    c_\infty^2 \Sigma_T \left(\frac{\Sigma}{\Sigma_T}\right)^\gamma  & (\Sigma > \Sigma_T); \\
  \end{array}
  \right.
\end{equation}
that is, the EOS changes from isothermal to adiabatic at a transition density $\Sigma_T$. This prescription approximates the structure of embedded protoplanets in disks \citep[see][]{InabaIkoma2003,OrmelKobayashi2012}. Thus, $\Sigma_T$ is a proxy of the thermodynamic state of the atmosphere, signifying the point where cooling becomes ineffective. Larger (grain) opacities and more luminous protoplanets will have a smaller isothermal layer, meaning a lower $\Sigma_T$.

The enthalpy, through the definition in \eq{enthalpy}, becomes (in units where $c_\infty=1$):
\begin{equation}
  W = \left\{ \begin{array}{ll}
    \displaystyle
    \log \Sigma & (\Sigma \le \Sigma_T) \\
    \displaystyle
    \frac{ \gamma}{\gamma-1} \left[ \left( \frac{\Sigma}{\Sigma_T} \right)^{\gamma-1} -1 \right] +\log \Sigma_T  & (\Sigma > \Sigma_T) \\
  \end{array}
  \right.
  \label{eq:W-expr}
\end{equation}
This equation is plotted as function of $\Sigma$ in \fg{wmodel} for several transition densities. In \eq{W-expr} the integration constants have been chosen such that $W$ is continuous at $\Sigma_T$ and $W(0)=0$. The adiabatic index is fixed at $\gamma=1.4$.

\section{A linear, analytic solution to the perturbed flow pattern}
\label{sec:linear}
\begin{figure}
  \centering
  \includegraphics[width=84mm]{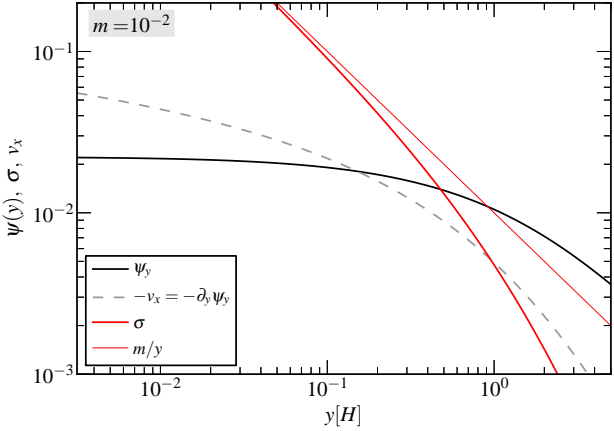}
  \caption{The perturbed stream function $\psi_y$ (black solid curve) and the radial velocity $-v_x$ as function of $y$ in the linear approximation of the linear model for a mass parameter of $m=10^{-2}$. The perturbed surface density which follows from \eq{sigma} is given by the red thick line. The red thin line gives $\sigma$ in the hydrostatic limit: $\sigma = m/r$.}
  \label{fig:hs}
\end{figure}
Assuming that the perturbations are small, \eqs{D-nondim}{B-nondim} can be linearized by setting $\Sigma = 1+\sigma$, $\Psi = \Psi_\infty + \psi$ where $\sigma$ and $\psi$ are assumed to be small with respect to the unperturbed quantities. Inserting these expressions into \eqs{D-nondim}{B-nondim} and keeping only the first order perturbations (that is, ignoring terms of order higher in $\psi$ and $\sigma$), one obtains (see \app{linear} for details): 
\begin{equation}
  \nabla^2 \psi +(\partial_x \sigma)({\cal M}_\mathrm{hw} -\omega_\infty x) +\sigma (\omega_\infty +\tilde\omega) = 0
  \label{eq:D-linear}
\end{equation}
\begin{equation}
  \sigma [1 -(\omega_\infty x -{\cal M}_\mathrm{hw})^2] +(\omega_\infty x -{\cal M}_\mathrm{hw}) (\partial_x\psi) = -\tilde\omega\psi +\frac{m}{r}
  \label{eq:B-linear}
\end{equation}
This set of equations, although simpler than the original, still represents a complex partial differential equation (PDE) which generally can be solved only numerically. To nevertheless pursue with an analytical model, two approximations are made:
\begin{enumerate}
  \item The perturbed flow depends only on $y$, $\psi=\psi(y)$;
  \item The limit $x\ll1$ and $x\ll y$ are considered. 
\end{enumerate}
The first approximation implies that the perturbed flow only describes the $x$ component of the velocity.  This is reasonable since the $y$-component of the flow is in any case dominated by the unperturbed solution ($\omega_\infty x\mathbf{e}_y$), whereas $v_x$ is entirely determined by the perturbed component. Thus, $\partial_x\psi=0$ and $v_x = (\partial_y \psi)/\Sigma \approx \partial_y \psi$ in the perturbed limit. The second assumption implies that the focus lies on the co-orbital region, $x\approx0$. This is again reasonable as $\Psi_\infty$ is small here. Since terms including $x$ are now small they can be neglected and \Eq{B-linear} becomes simply
\begin{equation}
  \label{eq:sigma}
  \sigma \approx -\tilde\omega\psi +\frac{m}{r},
\end{equation}
where we have also neglected the ${\cal M}_\mathrm{hw}^2$ term.

With the same argumentation, the middle term in \eq{D-linear} can be neglected; inserting \eq{sigma} into \eq{D-linear} and substituting $y$ for $r$ (in fact setting $x=0$) results in an ordinary differential equation (ODE) for $\psi(y)$:
\begin{equation}
  \label{eq:psi-gen}
  \partial_y^2 \psi -2\tilde\omega(1+\omega_\infty) \psi 
  = -\frac{2(1+\omega_\infty)}{r}m
\end{equation}
The solution to this ODE is denoted $\psi_{y0}$.  Closed-form solution exists, but depend strongly on the value of $\omega_\infty$. In a Keplerian disk, $\omega_\infty=-3/2$ it reads:
\begin{equation}
  \label{eq:psi0}
  \psi_{y0} =
  m\frac{ \left[\pi  |y|-2y \Si(\tilde{y})\right]\cos \tilde{y} 
    +2y \Ci(\tilde{y}) \sin \tilde{y}}{\sqrt{2} |y|}
\end{equation}
where $\tilde{y}=y/\sqrt{2}$ and $\mathrm{Ci}(u), \mathrm{Si}(u)$ the sine and cosine integral functions, defined as
\begin{equation}
  \mathrm{Ci}(u) = -\int_{u}^\infty \frac{\cos(t)}{t}dt; \qquad
  \mathrm{Si}(u) = \int_0^u \frac{\sin(t)}{t} dt.
  \label{eq:cisi}
\end{equation}
The general solution to \eq{psi-gen} contains two integration constants. These are of sinusoidal nature, $\psi \sim \sin y$, related to the solutions of the homogeneous equation. These oscillatory terms are discarded however, since such oscillatory terms do not decay for $y\rightarrow\infty$. Thus, \eq{psi0} only contains the particular solution to \eq{psi-gen}. Note that \eq{psi0} is linear in $m$. 

\begin{figure*}
  \centering
  \includegraphics[width=150mm]{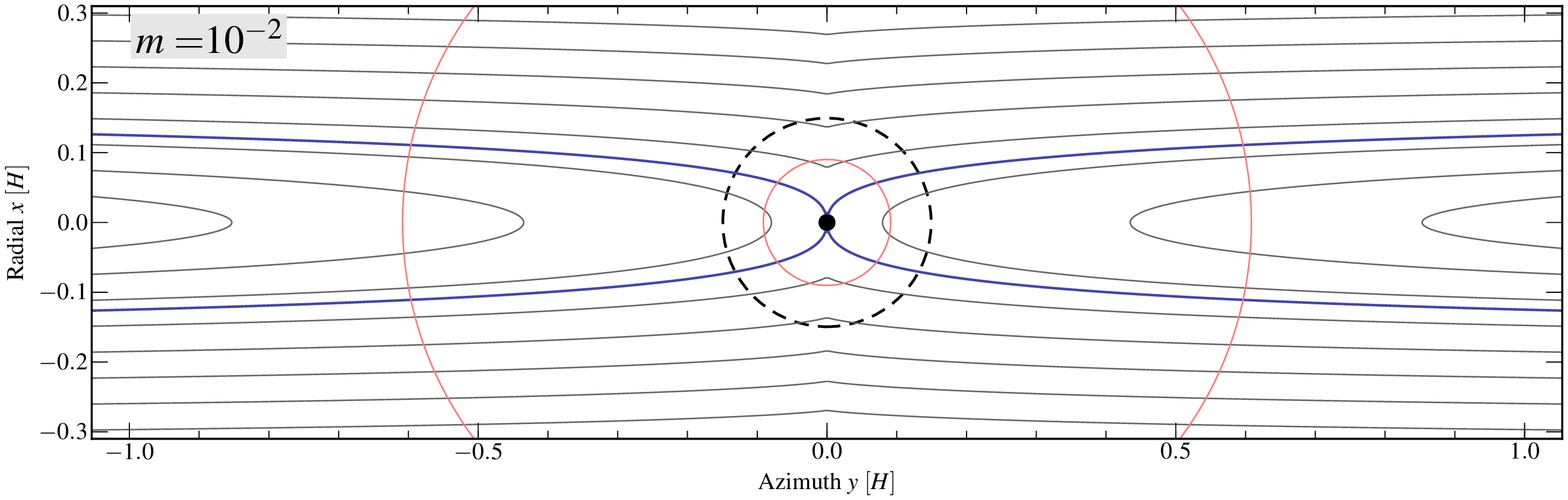}
  \caption{Flow pattern surrounding a protoplanet of dimensionless mass $m=10^{-2}$ in the linear regime, where perturbations are by construction small. Solid curves are contours of constant $\Psi = \Psi_\infty +\psi_{y0}$, where $\psi_{y0}$ is given by \eqp{psi0}. The far-field solution ($\Psi_\infty$) does not include a headwind term (shear-only case where $v_\mathrm{hw}=0$). Pink circles indicate the perturbation in surface density: contours of $\sigma=0.01$ and 0.1 are shown. The Hill radius ($r_h=(m/3)^{1/3}$) is given by the dashed circle and the Bondi radius ($r_b=m$) is indicated by the black dot at the centre. These solutions are used as boundary condition for the numerical calculations in \se{simul-res}.}
  \label{fig:hs-orbits}
\end{figure*}
\Eq{psi0}, plotted in \fg{hs}, peaks at $y=0$ for which $\psi_{y0}=\psi_\mathrm{hs}\equiv\pi m/\sqrt{2}$. With the caveat that $\sigma$ is no longer small here and that as a consequence the solution will break down when $y\rightarrow0$, this value of $\psi$ corresponds, in the shear-only case (${\cal M}_\mathrm{hw}=0$), to the widest streamline that undergoes so called horseshoe motion -- \ie\ where the gas elements upon their approach to the central object make a U-turn. This critical streamline therefore defines the width of the horseshoe region ($x_\mathrm{hs}$), which, as already alluded to in \se{intro}, is of special significance since it determines the horseshoe torque and connect to the planet's atmosphere. The horseshoe width $x_\mathrm{hs}$ is obtained by equating $\psi_\mathrm{hs}$ with the far-field solution, $\Psi_\infty(x_\mathrm{hs})$, \eq{psiinf}:
\begin{equation}
  \label{eq:hs-lin}
  x_\mathrm{hs} = \sqrt{ \frac{4\pi}{3\sqrt{2}} m } \approx 1.72 \sqrt{m}.
\end{equation}
($= 1.72 a_0\sqrt{q_p/h}$ in physical units; see \se{units}). \Eq{hs-lin} is in excellent agreement with the findings of \citet{PaardekooperPapaloizou2009} who found a prefactor of $1.68$.

\Fg{hs-orbits} shows the flow pattern in the linearized approximation with ${\cal M}_\mathrm{hw}=0$ and $m=10^{-2}$. Solid grey curves are iso-contours of $\Psi_\infty +\psi_{y0}$. The critical horseshoe streamline $\Psi_\mathrm{hs}$ is highlighted blue. Iso-density lines, which follow from \eq{sigma}, are in red. \Fg{hs-orbits} recovers the global features of the perturbed shear flow (the horseshoe orbits in particular) at distances far from the planet (\ie\ for $r\gg m$) as seen in many previous works \citep{MassetEtal2006,PaardekooperPapaloizou2009} . For large $x$ the linearized approximation which relies on $x\ll1$ will break down. For even larger distances, \ie\ for $|x|>2/3$, the PDE describing the system becomes hyperbolic, signifying that a (density) wave is excited. Hyperbolic (wave-like) equations require very different solution techniques \citep{GoldreichTremaine1980,Ward1986,TanakaEtal2002}. Including this feature is beyond the scope of this paper. 

For the purposes of this paper it is the flow structure in the vicinity of the perturber's Bondi radius that interests us. At these scales, the linear approximation is expected to break down and the nonlinear equations must be solved in earnest.  The analytical solution in the linear regime presented by \eq{psi0} will serve as the (outer) boundary conditions for the nonlinear calculation.

\section{Steady-state Flow simulations}
\label{sec:simul-res}
In this section a strategy to numerically solve the general equations for $\Sigma$ and $\Psi$ is discussed (\se{num-alg}). Then, a parameter study is presented, varying the mass of the perturber and the value of the headwind.

\subsection{Numerical algorithm}
\label{sec:num-alg}
\Eqs{D-nondim}{B-nondim} are solved by iteration, considering one equation at a time. First an initial $\Psi$ is assumed, for example the linear solution, $\Psi_\infty+\psi_{y0}$.  \Eq{B-nondim} is a scalar equation; given $\Psi(\bmath{x})$ and $\nabla\Psi$ it can be inverted to obtain $\Sigma$ at every grid point. Then, using these values for $\Sigma$, \eq{D-nondim} -- a diffusion equation -- is solved for $\Psi$. This completes one cycle. 

Thus, when considering \eq{B-nondim} the RHS is known. The LHS depends on the model for the enthalpy $W$, for which an equation of state (EOS) needs to be specified. 
Obtaining $\Sigma$ from \eq{B-nondim} is not without ambiguity, as is also illustrated in \fg{wmodel}. Here the LHS of \eq{B-nondim}, $f_\mathrm{lhs}(\Sigma)$, is plotted as function of $\Sigma$ for some arbitrarily-chosen value of $|\nabla\Psi|^2$. Starting from $\Sigma=0$, $f_\mathrm{lhs}(\Sigma)$ first decreases, reaches a minimum at $(\Sigma_\mathrm{min},f_\mathrm{min}$), and then increases towards infinity. Thus, for a given value of the RHS of \eq{B-nondim} there are either zero, one, or two solutions. In the case of two solutions, we always choose the largest $\Sigma$. This is simply because it gives the correct solution in the far field, provided no discontinuities (shocks) are present.

However, during the iteration process, cases are encountered in which the RHS of \eq{B-nondim} evaluates to less than $f_\mathrm{min}$. This is undesired. If this situation occurs, we re-adjust $\Psi$ by `mixing' it with the solution obtained in the previous iteration:
\begin{equation}
  \Psi = w_n \Psi_N +(1-w_n)\Psi_P,
\end{equation}
with $\Psi_P$ the previous solution for the stream function that did satisfy \eq{B-nondim}, $\Psi_N$ the new solution, and $w_n$ a weight. Starting from $w_n=1$, $w_n$ is adjusted until $\Psi$ satisfies \eq{B-nondim} again. In runs where $w_n$ in this way is forced to become arbitrary low the iteration procedure clearly has failed.  The calculations are then terminated.

The finite volume algorithm \texttt{FiPy} \citep{GuyerEtal2009},\footnote{Downloadable at http://www.ctcms.nist.gov/fipy/} is used to solve the partial differential \eq{D-nondim}. A grid and boundary conditions (BCs) need to be specified. It is customary for these kind of simulations to considered a rectangular grid. In this work, however, a polar grid ($r=\sqrt{x^2+y^2};\theta=\arctan(y/x)$) is used with the perturbing body located at the centre of the coordinate system. The inner boundary of the grid is located at a radius $r_1$ from the perturbing body and may correspond to the physical radius of the body. 

In \texttt{FiPy} either $\Psi$ or the gradient $\partial_r \Psi$ must be specified as a BC. At the inner boundary ($r=r_1$) the radial velocity, $v_r$, vanishes. Therefore, $\partial_\theta \Psi=0$ at the inner surface, and $\Psi$ is constant (say $\Psi=\Psi_1)$ on this surface. The value of $\Psi_1$ is \textit{a priori} unknown, but it can temporarily be set to an arbitrary value since \eq{D-nondim} will be unaffected by adding a constant to $\Psi$.  Therefore $\Psi_1$ is fixed at $\Psi_1=0$. The outer boundary ($r_\mathrm{out}$) is then specified by Neumann BCs, \ie\ $\partial_r \Psi(r_\mathrm{out}) = \partial_r (\Psi_\infty +\psi_{y0})$ where the far field solution in polar coordinates reads $\Psi_\infty = -\frac{1}{2}w_\infty r^2 \cos^2\theta +v_\mathrm{hw} r\cos\theta$ and $\psi_{y0}$ is given by \eq{psi0}.

The obtained solution (with $\Psi_1=0$) generally shows a mismatch at the outer boundary, \ie\ $\Psi(r_\mathrm{out}) \neq \Psi_\mathrm{out}$, where the latter is given by the linear perturbation: $\Psi_\mathrm{out} = \Psi_\infty(r_\mathrm{out}) +\psi_{y0}(r_\mathrm{out})$.  Let the difference between $\Psi$ and $\Psi_\mathrm{out}$ be denoted $C_\Psi$.  Although the constant $C_\Psi$ is irrelevant for the velocity of the flow (which are derivatives of $\Psi$), it matters for \eq{B-nondim}, since it was assumed in its derivation that $\Psi$ approaches $\Psi_\infty$ in the far field. The constant $C_\Psi$ is obtained by minimizing the relative error at $r=r_\mathrm{out}$,
\begin{equation}
  \sum_i \left( \frac{\Psi_i+C_\Psi -\Psi_{\mathrm{out},i}}{\Psi_{\mathrm{out},i}} \right)^2,
\end{equation}
with respect to $C_\Psi$, where the index $i$ refers to the $i^\mathrm{th}$ azimuthal point on the outer surface.  It is then added to the solution obtained by \texttt{FiPy} such that it approximately matches $\Psi_\mathrm{out}$ at $r_\mathrm{out}$. 

\begin{figure*}
  \centering
  \includegraphics[width=\textwidth]{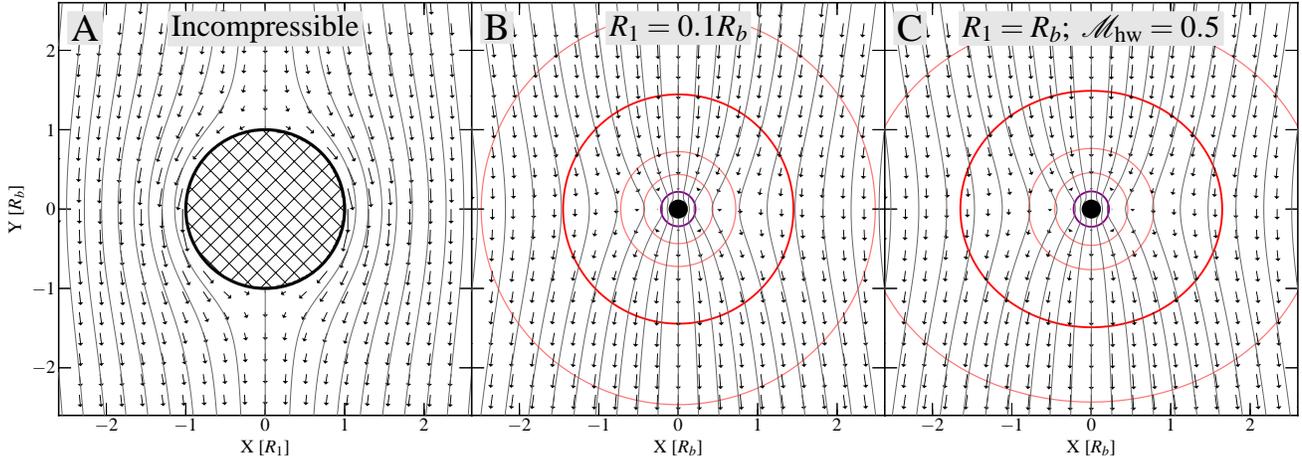}
  \caption{2D inviscid flow calculations in the case of a pure headwind (no terms involving shear or rotation). (a) Incompressible flow with the Mach number of the headwind equal to 0.1; (b) gravitating flow with the Bondi radius equal to one-tenth of radius of the object; (c) gravitating flow at an increased Mach number of 0.5. In each panel the flow pattern is described by arrows and streamlines (gray curves) that are isocontours of $\Psi$. There is no absolute scale for the size of the velocity arrows among the panels.  Solid, coloured contours indicate gas overdensities ($\Sigma/\Sigma_\infty$) of: 1.2, 1.5 (salmon), 2.0 (thick red), 4.0 (salmon), 10, $10^2$, $10^3$, \etc\ (all magenta). Note the focusing of streamlines starting in (b) and the distortion of the isodensity contours in (c).}
  \label{fig:headw}
\end{figure*}

\subsection{General results; key parameters}
\begin{table}
  \centering
  \begin{tabular}{llllll}
  \hline
  \hline
    Parameter         & Symbol                  & Default & Range         & Unit \\
  \hline
    Resolution        & $N_\mathrm{res}$        & 200           & 100---400     &\\
    Headwind          & ${\cal M}_\mathrm{hw}$  &               & 0, 0.05, 0.1  & $c_s$ \\
    Mass              & $m$                     & $10^{-2}$     & $10^{-3}$---$10^{-1}$  & $h^3 M_\star$ \\
    Inner radius      & $R_1$                   & 0.1           & $10^{-3}$---1 & $R_b$ \\
    Domain size       & $r_\mathrm{out}$        & 0.3           & 0.01---0.6    & $H$\\
    Transition mass   & $\Sigma_T$              & 0.1           & $10$---$10^3$ & $\Sigma_\infty$\\
  \hline
  \end{tabular}
  \caption{Default value and range of numerical and physical parameters considered in this study. The last column denotes the unit to convert the dimensionless numbers into physical ones. }
  \label{tab:pstud}
\end{table}
By virtue of the various simplifications that have been employed, the algorithm to solve for the steady-state flow pattern is computationally quite efficient; the CPU time is not much longer than a minute for a typical run on a standard desktop machine at the default resolution of 200 grid points in the radial and azimuthal direction. This allows us to carry out an extensive parameter study, studying the effects of the physical and parameter parameters see \Tb{pstud}. The key physical parameters of interest here are: the headwind ($v_\mathrm{hw}={\cal M}_\mathrm{hw}c_\infty$) and the planet mass (which is expressed in terms of the Hill radius $r_h$, see \se{units}). 

Not all runs converge. As expected from the discussion in the previous section, runs with a radial width that approaches the scaleheight no longer converge, because the setup does not resolve shocks. Either one of \eqs{D-nondim}{B-nondim} then fails to be satisfied. Typically, it is found that models with an outer radius ($r_\mathrm{out}$) exceeding $\simeq0.4$ start to slow down significantly (because $w_n$ becomes low) and models with $r_\mathrm{out}>0.6$ fail to converge at all. The dependence on the outer radius is therefore investigated, see \se{shear-only}. 

The sensitivity of the flow pattern to the number of grid points in the radial and azimuthal directions ($N_r, N_\theta$) has also been investigated. Generally, we find that convergence is achieved by $N_\mathrm{res}=200$, meaning that the $N_\mathrm{res}=400$ results did not provide a noticeably different result. But see \se{shear-only} for some exceptions.

The flow and density structure interior to the Bondi radius is affected by the inner radius $R_1$ and transition density $\Sigma_T$. These are also varied: $R_1$ starting from the Bondi radius until $R_h/10^3$ (which corresponds to the true physical radius at 5 AU); and $\Sigma_T$ is sampled at $10, 10^2$ and $10^3$.  The density close to $R=R_1$ can become rather large; and due to the conservation of vortensity ($\tilde\omega/\Sigma$), if non-zero, a very large azimuthal velocity emerges near the inner radius $R_1$. The large dynamic range in $\Sigma$ and $\bmath{v}$ in these cases often proves too much of a burden for the numerical code.
However, variations in $R_1$ and $\Sigma_T$ are found not to significantly affect the flow pattern outside the Bondi radius.

\subsection{Runs without shear or rotation (headwind-only)}
\label{sec:hw-only}
In this section all parameters regarding rotation and shear are zero: $\omega_\infty=\tilde\omega=\Omega=0$. Consequently, the outer outer boundary condition $\Psi_\mathrm{out}$ is set by the far field solution ($\Psi_\infty$) and does not involve $\psi_{y0}$, as the latter is only applicable for Keplerian shear.

\subsubsection{Incompressible runs}
When the flow is furthermore incompressible ($\Sigma=\Sigma_\infty=1$) \eq{D-nondim} becomes Laplace's equation, $\nabla^2 \Psi=0$. For a cylindrical geometry, the flow has an analytical solution \citep[\eg][]{LandauLifshitz1959}:
\begin{equation}
  \Psi 
  = {\cal M}_\mathrm{hw} \left(r -\frac{r_1^2}{r}\right) \cos \theta 
  = \Psi_\infty -\frac{ {\cal M}_\mathrm{hw} r_1^2 \cos\theta}{r},
  \label{eq:psi-cyl}
\end{equation}
where the unperturbed flow is assumed to move in the negative $y$ direction.  This solution can be used to test the accuracy of the numerical model. 

\Fg{headw}a shows the stream pattern corresponding to the incompressible limit, where $\Sigma$ is enforced to be unity. In the figure, the direction of the flow is indicated by black arrows, which indicate the direction of the flow and the flow velocity (larger arrows correspond to larger $v$ but there is no absolute scale in any of the figures). Contours of constant $\Psi$ -- streamlines -- lie parallel to $v$. Note the vertical streamline at $x=0$ ($\theta=\pi/2$) that must hit the objects due to symmetry considerations. The surface potential ($\Psi_1 = \Psi(r_1)$) is therefore characterized by this value for $\Psi$; here, $\Psi_1=0$.

The computed flow pattern corresponds very well to the analytical solution (\eqp{psi-cyl}), but it does not give a perfect match. The small offset in $\Psi$ is caused by the boundary conditions at the (finite) outer radius. Here (at $r=r_\mathrm{out}$) $\partial_r\Psi=\partial_r\Psi_\infty$, which deviates from \eq{psi-cyl} by a term of ${\cal O}(r_\mathrm{out}^{-2})$. By increasing $r_\mathrm{out}$ this error can be made arbitrarily small, however.

\subsubsection{Compressible, gravitating flow}
In \fg{headw}b gravity has been including while keeping the flow irrotational. The value of the headwind ($v_\mathrm{hw}$) now matters and it is fixed at 10\% of the sound speed. The flow pattern is further determined by the ratio of the Bondi radius to the inner radius, $R_1/R_b=0.1$. Isocontours of $\Sigma$ are nearly circular and displayed in \fg{headw}b,c in red-shaded colours: $\Sigma=$1.2, 1.5, 4 (salmon), 2 (red), $10, 10^2, 10^3$ (magenta).

For $R_b=R_1$ the flow does not significantly deviate from the incompressible limit. However, when the body increases in mass, it acquires a thick atmosphere. In the hydro\textit{static} limit $\Sigma(r)$ can be obtained by balancing the gravitational force to the pressure support:
\begin{equation}
  \label{eq:hydrostat}
  \frac{1}{\Sigma}\frac{dP}{dr} 
  = -\frac{m}{r^2},
\end{equation}
which results in $\Sigma = \Sigma_\infty \exp (m/r)$ for an isothermal EOS. The radius $r=m$ (which is the Bondi radius) signifies the point below which the gas density will increase exponentially. For $r_1<m$ density gradients becomes very steep, and huge amounts of gas mass can be concentrated towards $r=r_1$ as long as the EOS stays isothermal. Here however the EOS changes at a density $\Sigma_T=10^2$ (see \eqp{eos}), causing a transition to a power-law dependence of $\Sigma$ with $r$.

As can been seen in \fg{headw}b the (curl-free) flow focuses towards the gravitating body and slows down when it approaches the body. This is simply a consequence of mass conservation: as $\Sigma$ rises sharply, $v$ has to decrease due to the steady condition that is imposed on the flow. As a result, the flow converges towards the centre, and then diverges again. The focusing of the gas gives rise to an hourglass pattern for the streamlines.

For a Mach number of 0.1 in \fg{headw}b, the density is well-approximated by the hydrostatic limit of \solut{hydrostat}. Never the less the density isocontours are not entirely circular, but the deviation is unnoticeable by eye. For larger values of ${\cal M}_\mathrm{hw}$, however, the isocontours outside the Bondi sphere become more oval-shaped, see \fg{headw}c, where the Mach number of the incident flow is increased to 0.5. This trend agrees with the study of \citet{LeeStahler2011}, who considered an 3D axisymmetric geometry.  However, for cases where ${\cal M}_\mathrm{hw}\ll1$ the density can be well approximated as a function of radius only, $\Sigma = \Sigma(r)$. This finding will be employed in \se{approx} to find an analytical model for $\Psi$.

\begin{figure*}
  \centering
  \includegraphics[width=\textwidth]{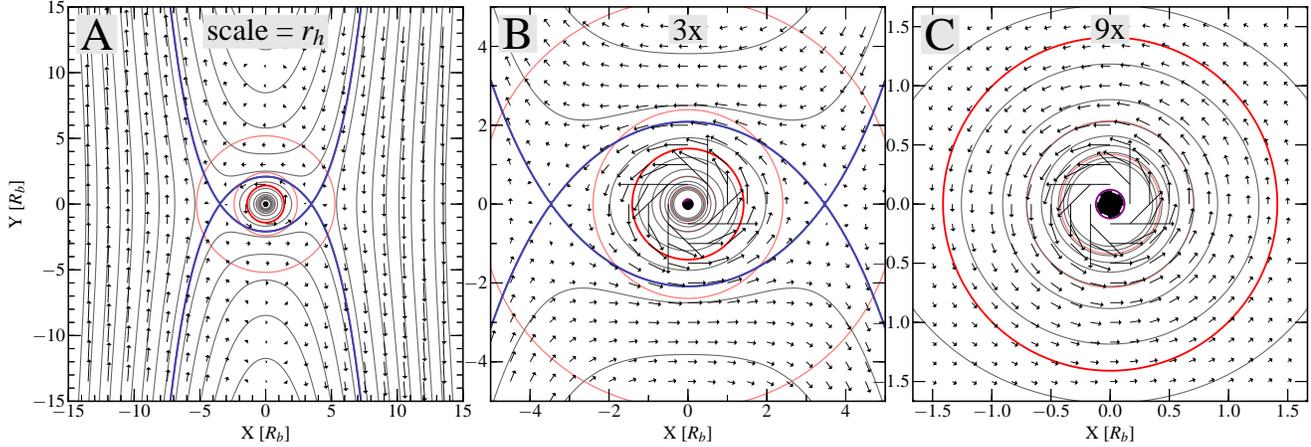}
  \caption{Flow pattern and density contours in a shear-only run (neglecting the headwind contribution) for $m=10^{-2}$ and an inner radius one-tenth of the Bondi radius. The panels give the flow pattern on scales of: (a) the Hill radius $r_h$; (b) $r_h/3$; (c) $r_h/9$ . See \fg{headw} for a description of the contours. In addition, the solid blue curves correspond to $\Psi_X$: the isocontour line of $\Psi$ that intersects the stagnation point of the flow on the $x$-axis (where velocities vanish). This streamline defines the horseshoe region (where streamlines make U-turns) and the atmosphere of the planet (where streamlines are closed). }
  \label{fig:shear}
\end{figure*}
\subsection{Shear-only runs; parameter study}
\label{sec:shear-only}
Next, the flow past bodies in a Keplerian potential is considered: \ie\ $\omega_\infty = -3/2$ and $\tilde\omega=1/2$ ($\Omega_0=1$) and the outer boundary condition at radius $r_\mathrm{out}$ involves $\psi_{y0}$ (\se{linear}). The headwind ${\cal M}_\mathrm{hw}=0$, such that the $y$-component of the flow velocity always vanishes at $x=0$. The mass is taken to be $m=10^{-2}$, which implies a Hill radius of $r_h=0.15$. In physical units, this mass equals $\sim$0.1--1.0$\mEarth$ depending on the disc aspect ratio (see \fg{dim-quants}).
The inner radius is again fixed at $r_1=0.1m$ ($R_1=0.1R_b$), which implies an inner boundary somewhat larger than the physical radius of the protoplanet. The larger value is adopted for computational reasons; the value of $r_1$ does not affect the flow pattern \textit{outside} the Bondi radius. 

In \fg{shear} the steady flow corresponding to these parameters is presented at three different magnifications. In \fg{shear}a the scale of the panel is that of the Hill radius, \ie\ the range in both $X$ and $Y$ is $2R_h$. For convenience, following \fg{headw}, length units on both axis are given in Bondi radii. \Fg{shear}b,c present zoom-ups of the flow in the vicinity of the planet.

The topology of the flow features several qualitatively distinct regions. To the far-left and right, at scales $|x|\gtrsim r_h$, it resembles the unperturbed solution: streamlines are nearly vertical with only little curvature. However, at smaller $x$-values the planet more strongly affects the flow, giving rise to the horseshoe region, quite similar to the linearized solution of \se{linear}. The streamline that divides these regions -- the separatix streamline $\Psi_X$ --  is highlighted. At the point where it crosses the $a$-axis the flow stagnates: $\bmath{v}=0$. The $x$-value of this stagnation point is nonzero, in contrast to the linear solution. A new region, not present in the linear solution, therefore arises: the planet's atmosphere.

\begin{figure}
  \centering
  \includegraphics[width=84mm]{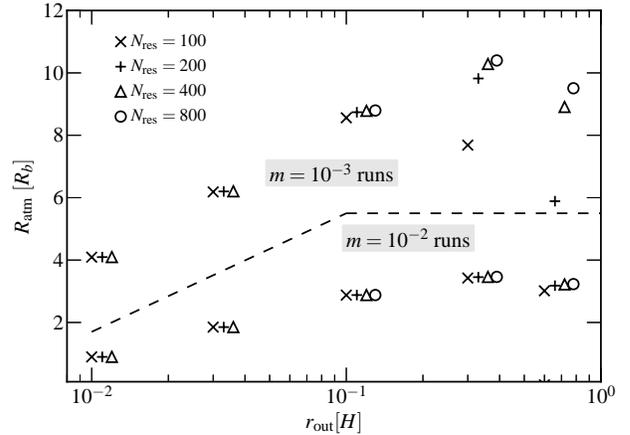}
  \caption{Dependence of the atmosphere size (the stagnation point where the critical streamlines intersect) $R_\mathrm{atm}$ (given in terms of Bondi radius) on the numerical resolution $N_\mathrm{res}$ (symbols) and the domain size $r_\mathrm{out}$ ($x$-axis). Symbols that corresponds to the same $r_\mathrm{out}$ are slightly offset for clarity. Results for the $m=10^{-3}$ runs lie above the dashed auxiliary line; runs with $m=10^{-2}$ below. This study indicates that convergence is achieved at a large domain size but that especially the low-$m$ runs require a large resolution.}
  \label{fig:resol}
\end{figure}
\label{sec:hs-width}
\begin{figure*}
  \centering
  \includegraphics[width=\textwidth]{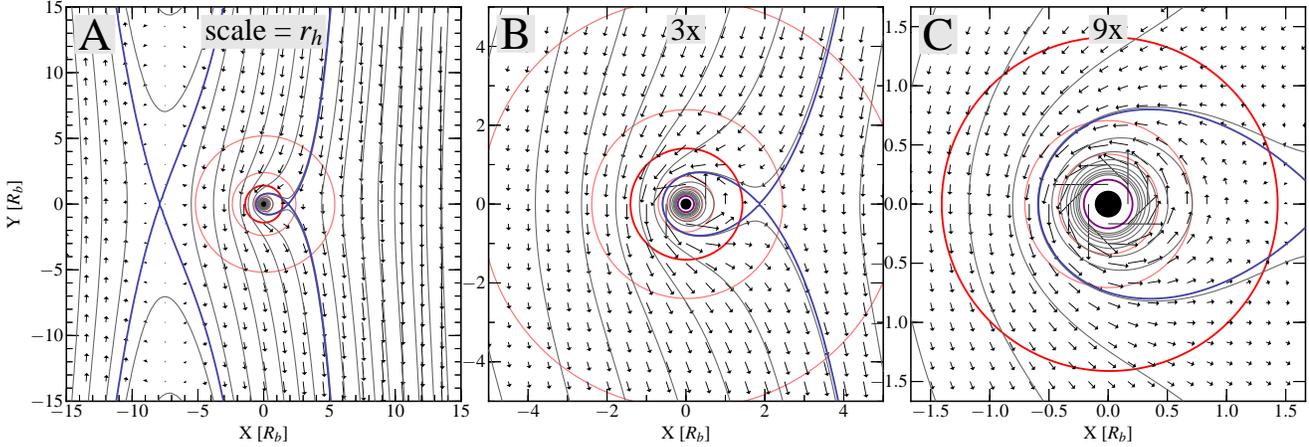}
  \caption{Same as \fg{shear} but also including a headwind of $v_\mathrm{hw}=0.1c_\infty$ (${\cal M}_\mathrm{hw}=0.1$). There are two solutions for the separatix streamline (in blue): one associated with the horseshoe region, which lies at $x\simeq-10$ in (a); and one circumscribing the planet's atmosphere. Note in (c) that the atmosphere is asymmetrical with nebula material penetrating within a fraction of the Bondi sphere.  See the caption of \fg{headw} for the description and labeling of the contour levels. }
  \label{fig:mixed}
\end{figure*}
Within the atmosphere region streamlines are closed and the gas, which encircles the planet, is bound.  The direction of the flow's rotation is prograde \citep[\eg][]{D'AngeloEtal2002,TanigawaEtal2012}. For the shear only runs, it encompasses a region that is quite larger than the Bondi radius, $r=m$. Most of the atmosphere's volume -- not necessarily its mass -- is therefore of low (nebular) density. At scales $r<r_b$ the density sharply increases. So does the velocity, which becomes more and more azimuthal ($|v_\phi/v_r| \gg 1$). This increase follows from vortensity conservation, \ie\ $\omega \simeq \tilde\omega\Sigma$. Nevertheless, the atmosphere at this stage is predominantly supported by pressure, instead of rotation.

These results are broadly in agreement with previous studies \citep[\eg][]{Miki1982,KorycanskyPapaloizou1996,BateEtal2003,MachidaEtal2010}, although usually a more massive planet is considered (with a non-zero softening radius instead of a surface) and the focus lies on resolving the flow pattern on scales $r\gg H$ in order to resolve the spiral wave pattern excited by the Lindblad torques. Nevertheless, the general features described above are all recovered.  The steady state solution of \citet{BateEtal2003} assumes that the planet accretes mass in, essentially, the BHL-regime, which may be appropriate for large planet masses, but requires a very efficient cooling mechanism in the low planet regime (\cf\ discussion in \se{intro}).


The radius of the atmosphere $R_\mathrm{atm}$ may be defined by the point where the separatix streamlines intersect, \ie\ it gives the distance to the stagnation point. Judging from \fg{shear} the atmosphere radius is $\approx$3.5 $R_b$. In \fg{resol} the dependence of $R_\mathrm{atm}$ against the two key numerical parameters -- the numerical resolution $N_\mathrm{res}$ and the domain size $r_\mathrm{out}$ -- is investigated both for an $m=10^{-2}$ and $m=10^{-3}$ perturber. The numerical resolution varies from $N_\mathrm{res}=100$ (these runs take a few seconds to complete on a desktop machine) to $N_\mathrm{res}=800$ (runs can approach one hour). The domain size $r_\mathrm{out}$ goes up to $r_\mathrm{out}=0.6$ scaleheights -- the maximum where the runs still converge (see \se{num-alg}). In this figure, the atmosphere radius is also measured in Bondi radii. Note that the $m=10^{-3}$ runs (which lie above the dashed auxiliary line) have a Bondi radius that is a factor ten smaller than the $m=10^{-2}$ runs. Thus, in physical units $R_\mathrm{atm}$ is (as expected) larger for larger $m$. 

\Fg{resol} shows that in most -- but not all -- of the runs the numerical resolution hardly affects $R_\mathrm{atm}$. More significant is the dependence on the domain size $r_\mathrm{out}$. \Fg{resol} shows that a larger domain size results in larger atmosphere radii. Clearly, the outer boundary constraints influence the flow pattern to some extent and to be assured of convergence $r_\mathrm{out}$ must be chosen as large as feasible. However, for the $m=10^{-3}$ runs the dynamic range, \ie\ the ratio of $r_\mathrm{out}$ to the Bondi radius ($m$) becomes very large in this way and a high resolution is needed to solve for the flow pattern at every point. But \fg{resol} does show that all models converge for large $r_\mathrm{out}$ and $N_\mathrm{res}$. In the following we will therefore stick with the standard of $r_\mathrm{out}=0.3$ and $N_\mathrm{res}=200$ for the numerical parameters for which we expect the error in $R_\mathrm{atm}$ to be at most $10\%$.

\subsection{Runs including shear and headwind}
\begin{figure*}
  \centering
  \includegraphics[width=\textwidth]{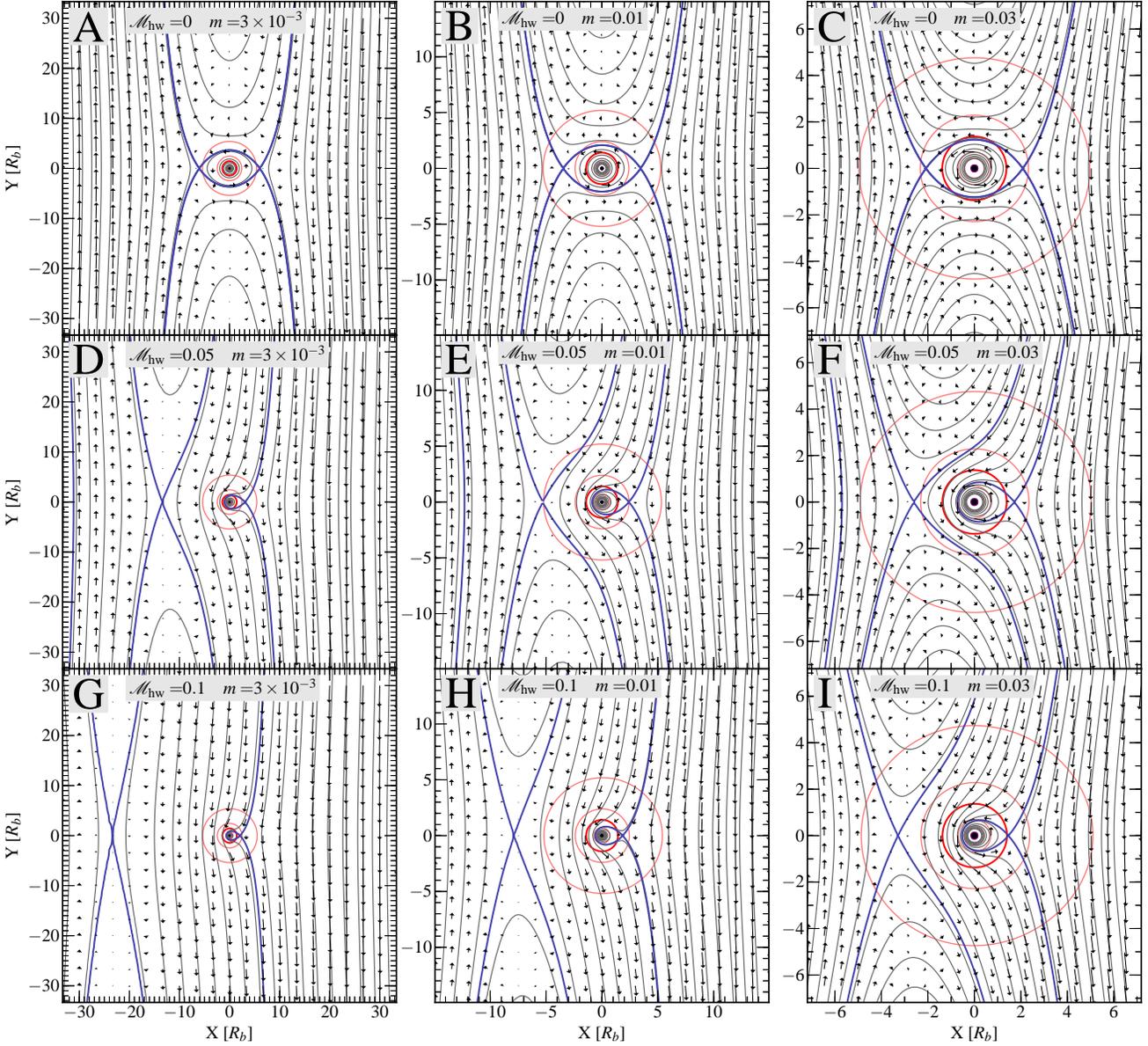}
  \caption{Trends with increasing planet mass (from left to right) and increasing  headwind (from top to bottom). Each panel covers a region two Hill radii in width, centred on the planet.}
  \label{fig:mixed3}
\end{figure*}
\Fg{mixed} shows the steady state flow pattern of a run that includes a nonzero headwind (${\cal M}_\mathrm{hw}=0.1$) and a (Keplerian) shear term, again for a mass of $m=10^{-2}$. The flow topology is quantitatively different from the shear-only case. The combination headwind+shear destroys the symmetry that was previously present along the $x=0$ axis; the stagnation points are no longer symmetrical around $x=0$. As a result, there are two separatix streamlines and two solutions for $\Psi_X$: one that encloses the planet's atmosphere ($\Psi_\mathrm{atm}$); and one, offset from the planet, that corresponds to the horseshoe region ($\Psi_\mathrm{hs}$). The planet's atmosphere is sandwiched by flow which moves in the direction of negative $y$ -- an effect that is also seen for lower $v_\mathrm{hw}$. This topology is therefore the rule and the $v_\mathrm{hw}=0$ run of \fg{shear} a special symmetric limit.

The position of the separatix streamlines changes according to the planet mass and the value of the headwind. \Fg{mixed3} investigates the effects of varying the headwind (${\cal M}_\mathrm{hw}=0,\ 0.05$ and $0.1$; from top to bottom) and the planet mass ($m=3\times10^{-3},\ 10^{-2}$ and $3\times10^{-2}$; from left to right). Thus, \fg{mixed3}a corresponds to \fg{shear} and \fg{mixed3}d to \fg{mixed}. The scale of each panel is again the Hill radius.
Note that the left-most highlighted streamline in the runs with a modest headwind (${\cal M}_\mathrm{hw}=0.05$; panels d--f) is not associated with any stagnation point. Instead, it just happens to have the same value as that of the atmosphere streamline $\Psi_\mathrm{atm}$ that encloses the atmosphere of the planet. 

Despite its importance, many works studying the flow behavior tend to neglect the headwind term (but see, \eg\ \citealt{Paardekooper2009} for the implications of a large headwind on planet migration). In particular, the headwind may affect the accretion efficiency of small particles. Comparing the ${\cal M}_\mathrm{hw}\neq0$ to the ${\cal M}_\mathrm{hw}=0$ runs, it is noted that in the former the atmosphere region is much reduced. Small particles can be brought closer to the planet, even within a fraction of the Bondi radius. From this observation it seems that a (modest) headwind may actually promote protoplanet growth. \Se{p-traj} assesses this issue in more detail.

The asymmetry of the atmosphere is quantified in \fg{atm}, which plots the radius of the atmosphere as function of the mass of the planet for runs with and without a headwind. The size of the atmosphere $R_\mathrm{atm}^+$ ($R_\mathrm{atm}^-$) is defined to be the absolute value of the $x$-coordinate where $\Psi_\mathrm{atm}$ intersects the positive (negative) $x$-axis.  For ${\cal M}_\mathrm{hw}=0$, $R_\mathrm{atm}^+=R_\mathrm{atm}^-$ is seen to follow a trend that lies between the Bondi radius and the Hill radius. Thus, for a small planet it is significantly larger than the Bondi radius (\cf\ \fg{shear}). For ${\cal M}_\mathrm{hw}\neq0$ the symmetry breaking causes two solutions with $R_\mathrm{atm}^+ > R_\mathrm{atm}^-$. The negative solution always lies within the Bondi radius. In addition, $R_\mathrm{atm}^-$ and $R_\mathrm{atm}^+$ decrease with increasing headwind.

\section{Analytical description of the flow pattern near the perturber}
\label{sec:approx}
From the numerical results, it was found that the density $\Sigma$ is primarily a function of $r$. This can be understood from comparing the terms appearing in \eq{B-nondim}. When $r\lesssim m$ the term representing the planet's potential, $m/r$, is dominant, whereas for large radius the first order solution of \eq{B-nondim} is, by definition, $\Psi_\infty$. At $r\sim m$, then, the Bernoulli equation approximately reduces to $W(\Sigma)=m/r$, which in the isothermal regime leads to $\Sigma = \exp(m/r)$.

\begin{figure}
  \centering
  \includegraphics[width=84mm]{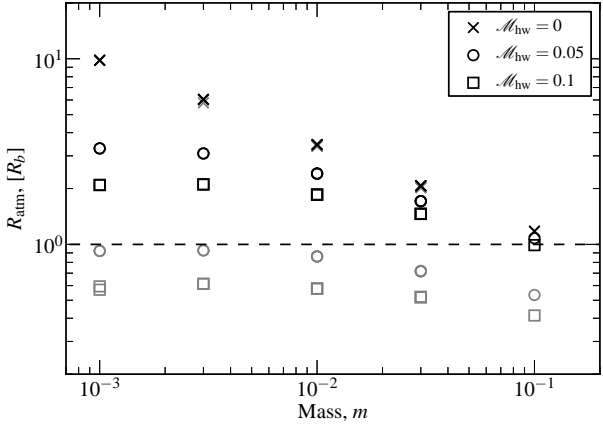}
  \caption{Atmosphere size as defined by the distance to the stagnation points. The solution corresponding to the positive $x$-axis ($R_\mathrm{atm}^+$) is plotted by black symbols, while gray symbols correspond to $R_\mathrm{atm}^-$.  In the shear-only case (${\cal M}_\mathrm{hw}=0$) symmetry ensures that $R_\mathrm{atm}^+=R_\mathrm{atm}^-$. A headwind breaks the symmetry and results in a noticeable decrease, of especially $R_\mathrm{atm}^-$.}
  \label{fig:atm}
\end{figure}
The solution $\Sigma = \exp(m/r)$ corresponds to an isothermal atmosphere in hydrostatic equilibrium. In the following, the hydrostatic solution is supposed to hold for the density structure, but not for the flow itself which is steady, but not static. The diffusion \eq{D-nondim} is then left to solve. It can be written as
\begin{equation}
  -\nabla \log \Sigma \cdot \nabla\Psi +\nabla^2 \Psi 
  = -\Sigma ( \tilde\omega\Sigma -2 )
\end{equation}
or in terms of the derivatives towards $r$ and $\theta$:
\begin{equation}
  r^2\partial_{rr} \Psi +\left(r-r^2\frac{d\log\Sigma}{dr}\right)\partial_r \Psi +\partial_{\theta\theta}\Psi
  = -\Sigma r^2 (\tilde\omega\Sigma -2 ),
\end{equation}
when the density is a function of radius only. The term in brackets on the LHS becomes $(r+m)\partial_r\Psi$ for $\Sigma = \exp(m/r)$.

This partial differential equation (PDE) can be solved by adopting the \textit{Ansatz} that $\Psi$ is a superposition of a homogeneous solution, $\Psi_\mathrm{hom}(r,\theta)$, and a particular solution $\Psi_r(r)$:
\begin{equation}
  \label{eq:Psi-gen}
  \Psi(r,\theta) \equiv \Psi_\mathrm{hom}(r,\theta) + \Psi_r(r);
\end{equation}
\ie\ where $\Psi_\mathrm{hom}$ satisfies the homogeneous PDE
\begin{equation}
  \label{eq:hom}
  r^2\partial_{rr} \Psi_\mathrm{hom} +(m+r)\partial_r \Psi_\mathrm{hom} +\partial_{\theta\theta}\Psi_\mathrm{hom} = 0,
\end{equation}
and $\Psi_r(r)$ the ordinary differential equation (ODE)
\begin{equation}
  \label{eq:ihm}
  r^2\Psi_r'' +(m+r)\Psi_r'
  = e^{m/r} r^2 \left[ 2 -\tilde\omega e^{m/r} \right],
\end{equation}
where primes denote derivatives to $r$. 

This simplifies the problem greatly. \Eq{hom} resembles Laplace equation but contains an additional $\partial_r \Psi$ term. \Eq{ihm} only contains $r$ as the variable. Closed form solutions are possible.

\subsection{Solving \eq{hom}}
The strategy to solve this PDE is the same as with Laplace's equation. Assuming that the solution is separable, $\Psi_\mathrm{hom}(r,\theta) = f(r)g(\theta)$, \eq{hom} reduces to two ODEs:
\begin{equation}
  r^2 f'' +(m+r)f' -n^2 f= 0
  \label{eq:radial}
\end{equation}
for the radial part; and
\begin{equation}
  \frac{d^2 g}{d\theta^2} +n^2 g = 0
  \label{eq:azi}
\end{equation}
for the angular part. Thus, $g(\theta) = \cos n\theta$, where the sine solutions are discarded because of symmetry considerations -- the far field solution $\Psi_\infty$ only contains cosine terms.  Clearly, $n=\pm1$ correspond to the case of uniform flow (headwind) and $n=\pm2$ holds for the shear case. For $|n|\le2$ the solutions to \eq{radial}, denoted $f_n(r)$, read:
\begin{equation}
  f_{n}(r) = \left\{ \begin{array}{ll}
    \displaystyle
    12\left[e^{m/r}\left(\frac{2r}{m} -\frac{6r^2}{m^2}\right) 
    +1 +\frac{4r}{m} +\frac{6r^2}{m^2} \right] & (n=-2)\\[3mm] \displaystyle
    \frac{2r}{m}e^{m/r} -\frac{r}{m} -1       & (n=-1) \\[3mm] \displaystyle
        1                                     & (n=0) \\[3mm] \displaystyle
        1 +\frac{r}{m}                        &(n=1)  \\[3mm] \displaystyle
        \frac{1}{6} +\frac{2r}{3m} +\frac{r^2}{m^2}           & (n=2)
  \end{array} \right.
\end{equation}
For $r\gg m$, $f_n(r)$ is of ${\cal O}( [r/m]^n)$, also for negative $n$. The general solution for $\Psi_\mathrm{hom}$ therefore reads
\begin{equation}
  \label{eq:psihom}
  \Psi_\mathrm{hom}(r,\theta) = \sum_{n=-2}^2 c_n f_n(r) \cos n\theta,
\end{equation}
with $c_n$ a constant that must be obtained from the boundary conditions. 

\subsection{Solving \eq{ihm}}
\Eq{ihm} is only of first order in the derivative of $\Psi_r$, $\Psi_r'$, and exhibits an analytical solution. For $\Psi'(r)$ it reads 
\begin{equation}
  \label{eq:Dpsi-r}
  \Psi_r'(r) =
  \frac{e^{\frac{m}{r}} \left[ 
      m^2\tilde\omega\text{Ei}\left(\frac{m}{r}\right)
      - \tilde\omega r(m+r)  e^{\frac{m}{r}}
      + 2(r^2 +C_1) 
      \right]}{2 r},
\end{equation}
where $\Ei(x)$ is the exponential integral (see \eqp{expEi}) and $C_1$ a constant of integration. This constant determines the direction of rotation (prograde or retrograde) of the gas flow within the atmosphere. \Eq{Dpsi-r} can be integrated once more to provide $\Psi_r$ (see \app{full-sol}), which introduces another constant of integration, $C_2$.

\begin{figure*}
  \centering
  \includegraphics[width=\textwidth]{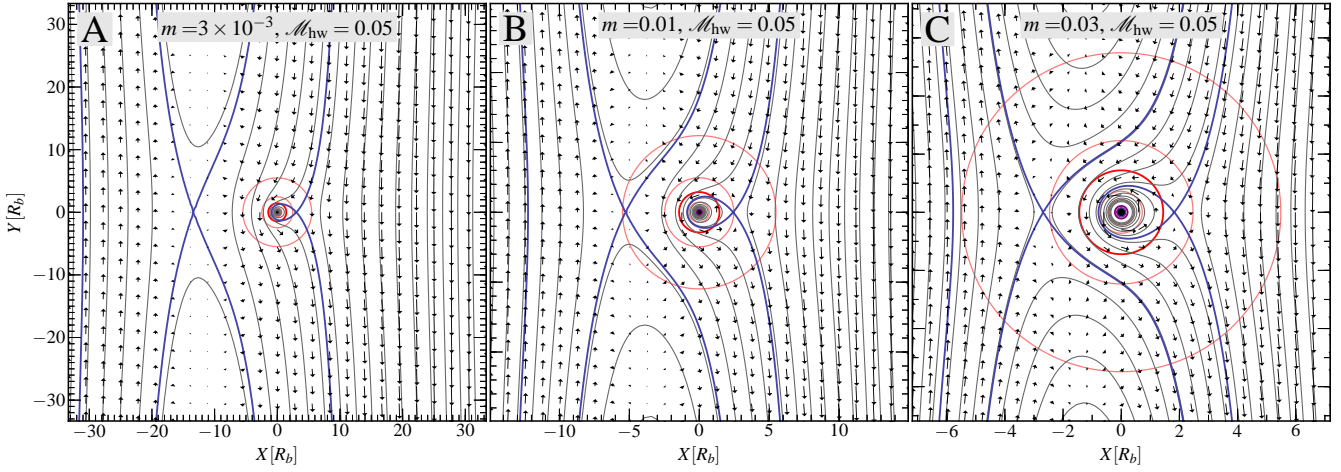}
  \caption{Flow patterns drawn from the analytical prescription for $\Psi$ outlined in \se{approx} for perturber masses of: (a) $m=3\times10^{-3}$, (b) $m=10^{-2}$, and (c) $m=3\times10^{-2}$. The headwind is ${\cal M}=0.05$ in each panel and the control parameter $y_c$ is set at 0.05 as argued in the text. These panels should be compared to the middle row of \fg{mixed3}.}
  \label{fig:synthetic}
\end{figure*}
\subsection{Setting the integration constants}
We first consider the limit $r\gg m$, where the solution $\Psi_\mathrm{hom} +\Psi_r$ should converge to the far-field solution, $\Psi_\infty$.  In the limit $r\gg m$, \eq{Dpsi-r} can be approximated as
\begin{equation}
  \Psi_r'(r) \approx \left( 1 -\frac{\tilde\omega}{2} \right)r +\left( 1 -\frac{3}{2}\tilde\omega \right)m,
  \label{eq:Dpsi-liminf}
\end{equation}
independent of $C_1$. Integrating once more:
\begin{equation}
  \label{eq:psir}
  \Psi(r) 
  \approx 
  -\frac{1}{4}\omega_\infty r^2 
  +\left( 1-\frac{3\tilde\omega}{2} \right)mr +\dots
\end{equation}
where $\omega_\infty$ has been used instead of $\tilde\omega$ in the first term.  Only terms of ${\cal O}(r^1)$ or higher are included in the expansion. For $r\gg m$, the linear term is small and will be neglected. Similarly, the expansion of $\Psi_\mathrm{hom}$ gives
\begin{equation}
  \Psi_\mathrm{hom} 
  \approx 
    c_2 \frac{r^2}{m^2} \cos2\theta
    -c_1\frac{r}{m}\cos\theta 
\end{equation}
It is now required that the solution $\Psi_\mathrm{hom} +\Psi_r$ matches the far-field solution $\Psi_\infty$ (\eqp{psiinf}), which can be written as
\begin{equation}
  \label{eq:psiinf-2}
  \Psi_\infty = -\frac{\omega_\infty r^2}{4}\cos 2\theta +v_\mathrm{hw}r\cos\theta -\frac{\omega_\infty r^2}{4}.
\end{equation}
The three terms correspond respectively to the $n=2$ (shear) term of the homogeneous solution, the $n=1$ (headwind) term, and the leading term of \eq{psir}.  Thus the integration constants of the homogeneous solution, $c_1$ and $c_2$, simply follow from the requirement that the leading terms should match $\Psi_\infty$ in the far field:
\begin{eqnarray}
  \label{eq:c-pos}
  c_2 &=&  -\frac{\omega_\infty m^2}{4}; \\
  c_1 &=&  m v_\mathrm{hw}.
\end{eqnarray}

The $c_{-1}$ and $c_{-2}$ constants can be found by demanding that $\Psi$ at the inner boundary ($r=r_1$) is a constant, \ie\ no azimuthal dependence. This gives
\begin{eqnarray}
  c_{-2} &=& -c_2 \frac{f_2(r_1)}{f_{-2}(r_1)}; \\
  c_{-1} &=& -c_1 \frac{f_1(r_1)}{f_{-1}(r_1)}. 
  \label{eq:c-neg}
\end{eqnarray}
But these terms do not affect the solution for $r\simeq m$ and are rather unimportant. One may neglect these and the negative $n$ terms of \eq{psihom}.

Much more critical are the integration constants of $\Psi_r$, $C_1$ and $C_2$. There is, unfortunately, no clear constraint that allows a straightforward determination. Somewhat more loosely, we can argue that the solution for $\Psi_r+\Psi_\mathrm{hom}$ should match to the solution in the linear regime, $\Psi_\infty+\psi_{y0}$, at some point. In particular, let us consider the horseshoe region for $x=0$. Thus, we consider the $y$-axis ($\theta=\pi/2$) and require that at a certain point $(x,y)=(0,y_c)$ $\Psi_\mathrm{hom} +\psi(y_c) = \psi_{y0} +\Psi_\infty$:
\begin{equation}
  \label{eq:psi-match}
  -c_2\frac{y_c^2}{m^2} + \Psi_r(y_c) = \psi_{y0}(y_c).
\end{equation}
(Note that $\Psi_\infty=0$ for $x=0$ and $\cos 2\pi/2=-1$.) In addition, it is required that the radial derivatives of the stream function match at this point:
\begin{equation}
  \label{eq:Dpsi-match}
  -2c_2\frac{y_c}{m^2} + (\partial_r\Psi_r)(y_c) = \partial_y\psi_{y0}(y_c).
\end{equation}
\Eq{Dpsi-match} will determine $C_1$, whereas \eq{psi-match} determines $C_2$. The integration constants follow straightforwardly from the $\Psi_r$ and $\Psi_r'$ expressions, although the very expressions are rather cumbersome. Assuming that $m\ll y_c \ll 1$ simplifies these significantly, however, and we may approximate:
\begin{align}
  \label{eq:Ci}
  C_1 \approx&\ m y_c \left[ \log \left(\frac{y_c}{\sqrt{2}}\right) +\gamma \right] \\
  C_2 \approx&\
  C_1 \left(\frac{m}{y_c}+\gamma \right)+\log \left(y_c\right) \left(m y_c-C_1\right) \\
   & +\frac{1}{2} m \left[(-2+2 \gamma -\log (2)) y_c+\sqrt{2} \pi \right]+C_1 \log (m),
   \nonumber
\end{align}
where $\gamma$ is the Euler-Mascheroni constant.

 The choice for the integration constants has now been rewritten in terms of $y_c$. Although there is no natural choice for $y_c$, it is expected that the hydrostatic assumption for $\rho$ will become less accurate for larger $r$. Judging from \fg{hs}, a typical scale will be $y\simeq0.1$. Thus, $y_c$ may represent the point where the nonlinear solution transfers to the linear solution of \se{linear}. Nevertheless, there is some freedom allowed in choosing the precise value of $y_c$. We find that with $y_c=0.05$ the solution very well matches those of the numerical simulations and we stick to this value.

In summary, we have derived an analytical approximation of the perturbed flow pattern for a low mass planet ($m\ll1$) under the assumptions that the density is radial, consistent with the isothermal EOS, $\Sigma=\exp(m/r)$, and that its solution reaches that of the unperturbed flow at $r\gg m$. The perturbed stream function is \eq{Psi-gen} with \eq{psir} for $\psi(r)$ and \eq{psihom} for $\Psi_\mathrm{hom}$ with integration constants given by \eqsto{c-pos}{Ci}.




\subsection{Synthetic streamline patterns}
\begin{figure*}
  \centering
  \includegraphics[width=\textwidth]{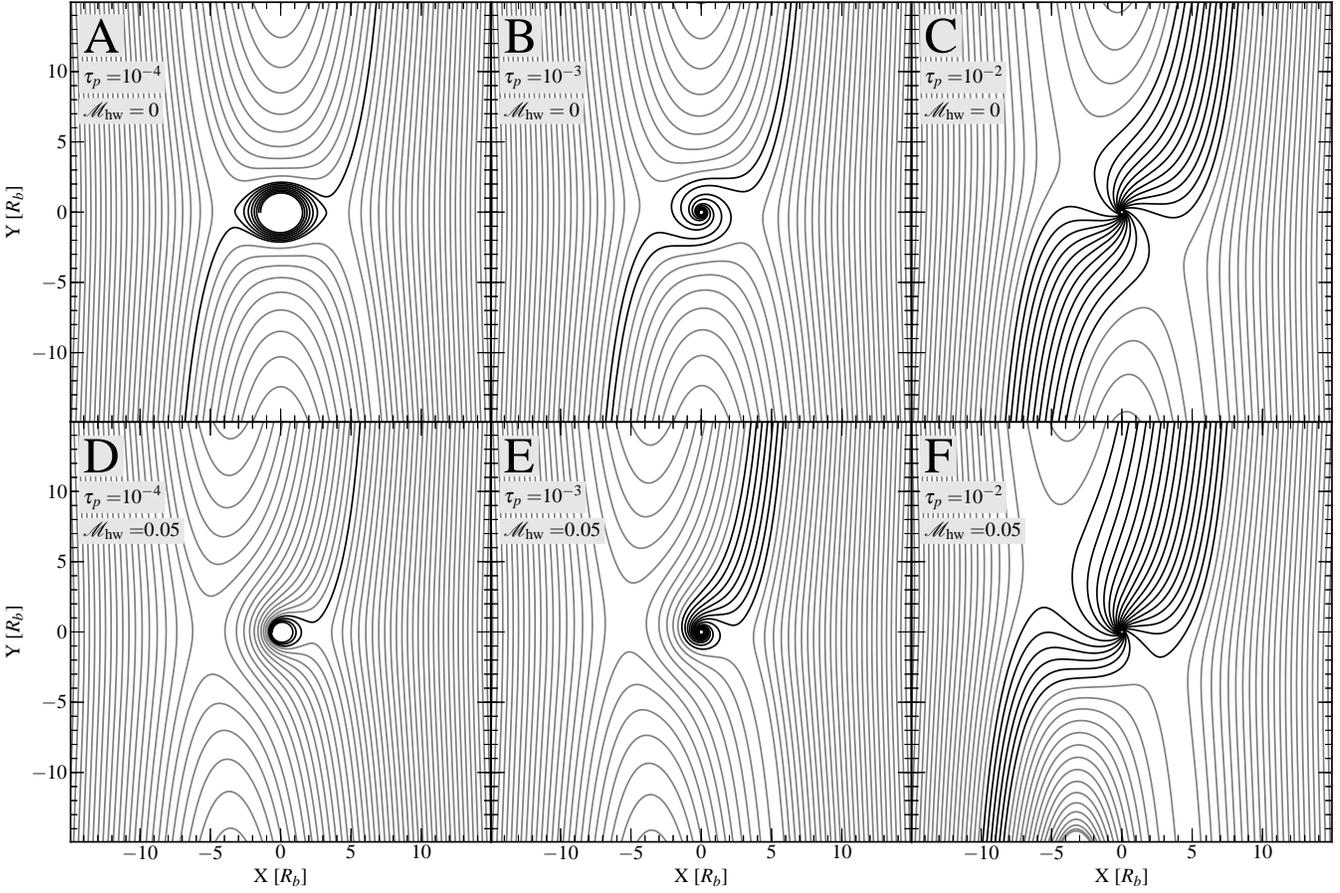}
  \caption{Trajectories of small, solid particles, characterized by their dimensionless friction time $\tau_p$, in the vicinity of an $m=10^{-2}$ planet. For $\tau_p\ll1$, particles largely follow the streamlines of the gas. Trajectories that result in accretion by the protoplanet are highlighted black. Top panels: ${\cal M}_\mathrm{hw}=0$; bottom panels: ${\cal M}_\mathrm{hw}=0.05$.}
  \label{fig:p-traj}
\end{figure*}
\Fg{synthetic} presents the streamline patterns derived from the analytical expressions presented above for the values of the mass parameter $m$.  The free parameter $y_c$ was fixed at 0.05. This figure can be compared to the middle panels of \fg{mixed3}, the corresponding numerical solutions. The analytic solution retrieves most features very well. The analytical description is advantageous, because it is much faster than running a full numerical simulation and does not require interpolation of the result.

\section{Trajectories of small particles and their accretion behavior}
\label{sec:p-traj}
Having quantified the velocity structure around the planet in the form of the analytic solution presented in \se{approx}, trajectories of solid particles can be calculated. Particle trajectories will deviate from those of the gas (the streamlines) due to their inertia. That is, it takes a solid particle a characteristic time -- the friction or stopping time $t_p$ -- to align its motion with that of the gas. The dimensionless equivalent to $t_p$ is $\tau_p = t_p\Omega_0$.  Big particles ($\tau_p\gg1$) like km-size planetesimals experience the gas only in the form of a small perturbation, which none the less is important to damp the random motions (eccentricity) of these bodies. Here, we consider small particles of $\tau_p\ll1$. To first approximation, they tend to follow the gas. However, their not fully perfect coupling to the gas causes them to slowly settle due to the gravitational force, \eg\ towards the central star \citep{Weidenschilling1977} or to planets \citep{OrmelKlahr2010}.

The particle size $\tau_p$ adds another dimension to the parameter space (along with $m$ and ${\cal M}_\mathrm{hw}$) and it is left to a follow-up work to conduct a full exploration. \Fg{p-traj} only presents a preview, where particle trajectories -- \textit{not} gas streamlines -- are plotted for $m=10^{-2}$, ${\cal M}_\mathrm{hw}=0$ and $0.05$, and $\tau_p=10^{-4}, 10^{-3}, 10^{-2}$. These correspond to small particles in the $\sim$100 $\mu$m--10 mm size range, dependent on the values of the local gas density, disc radius, and particle internal density\footnote{The relation between the size $s$ and friction time $t_p$ depends on the gas drag law. In the Epstein regime (small particle, low gas density), $t_p = \rho_s s/\rho c_\infty$ where $s$ is the particle size. When particle sizes become larger than the mean-free-path $l_\mathrm{mfp}$ of the gas, $t_p$ increases by a factor $\sim s/l_\mathrm{mfp}$. For even larger particle sizes $t_p$ starts to depend on the particle-gas velocity itself; but this regime is not considered here.}.

In \fg{p-traj} each curve represents a trajectory of a single particle. We have omitted drawing arrows to indicate their directions; but these are clear from comparison to \fg{mixed3}b and e. Highlighted curves indicate particles that are accreted by the protoplanet. They all settle to the planet due to inspiraling motion in the prograde manner. This means that the accretion rate only depends on the mass of the planet, not on its radius. This accretion mechanism qualitatively differs from the way planetesimals are accreted, where one relies on a sufficiently close encounter to `hit' the target. Accretion through settling may provide a very fast channel for growth. For $\tau_p\sim1$ particles it has already been shown that their accretion radii are on the order of the Hill radius \citep{OrmelKlahr2010,LambrechtsJohansen2012}. However, for smaller particles accretion cross sections are likely smaller since they couple more strongly to the gas. 

Indeed, \fg{p-traj}a,d shows that the $\tau_p=10^{-4}$ particles are difficult to accrete since their trajectories are strongly tied to the gas. The particle trajectories in these figures strongly resemble the gas streamlines of \fg{mixed3}. In both \fg{p-traj}a,d only two particle streams penetrate the planet's atmosphere and settle slowly towards the center (at some point when it is clear that the particle will be accreted, the integration is stopped). Most particles that approach the planet, however, cannot break through the atmosphere as they, being strongly coupled to the gas, are dragged along gas streamlines. 

\citet{PeretsMurray-Clay2011} introduced the wind-shearing radius $R_\mathrm{ws}$ to quantify the competing effects of the (central) gravitational two body force and the gas drag force (which pulls particle away from accretion). That is, only particles with impact parameters less than $R_\mathrm{ws}$ can be accreted. A similar wind-shearing radius can be defined for low-mass planets surrounded by an atmosphere. First, it is noted that only particles near the atmosphere boundary will be accreted. Therefore, particles traveling along the $\Psi_\mathrm{atm}$ streamline are most likely to be accreted. The wind-shearing radius then becomes the bandwidth $\Delta R_\mathrm{ws}$ over which the gravitational pull of the planet is sufficient to drag them into the atmosphere region. This idea will be explored further in a follow-up work.

Since the drag force equals $\mathbf{F} = (\mathbf{v}_\mathrm{p}-\mathbf{v}_\mathrm{gas})/t_p$, its influence will decrease with increasing friction time.  
For particles of friction time $\tau_p=10^{-3}$ (the middle panels of \fg{p-traj}) more particles accrete, but more so for the headwind case (7 counts) than the ${\cal M}_\mathrm{hw}=0$ case (4).  Even for particle streams that do not accrete, deviations from the gas streamlines can be noted in \fg{p-traj}b and e. For the $\tau_p=10^{-2}$ particles in \fg{p-traj}c and f the deviation from the streamlines is much more evident; close to the planet these particles decouple from the gas and the impact parameter for accretion is very large. Note finally that due to the prograde rotation of the gas in the atmosphere, small particles that enter the atmosphere will also start to rotate in a prograde manner; and once they have settled towards the central object they will convey this angular momentum to the accreting body, causing it to spin up. Small-particle accretion therefore naturally explains the observed preference for prograde spins of solar system bodies \citep{JohansenLacerda2010}.


Obviously, \fg{p-traj} only presents a small subset of the parameter space and the above argumentation is rather qualitative.  In a future work we will return to the question of the efficiency of small particle accretion and a provide a more quantitative analysis.

\section{Discussion}
\label{sec:discussion}
\subsection{Caveats and extensions of the numerical method}
We have described a method to compute the gas flow around gravitating bodies in steady state. The idea of the adopted approach is to solve for the stream function $\Psi$ -- a scalar quantity which fully describes the flow in 2D. The global flow pattern agrees well with previous numerical studies.  However, the method employs several idealizations (the flow must be inviscid, 2D, barotropic, subsonic, \etc), and it would be interesting to investigate whether these conditions can be relaxed. We remark that a supersonic expansion of the method has already been presented by \citet{KorycanskyPapaloizou1996}, which would allow us to explore particle accretion for more massive planets.

Since the focus of this work lies on embedded planets, an extension to three dimensions is perhaps the most obvious avenue. Indeed, 3D effects could have some effects, \eg\ with regard to the width of the horseshoe region \citep{BateEtal2003}.  Although the stream function formulation can be easily generalized to axisymmetric configurations \citep{LeeStahler2011}, it is not commonly used for truly 3D geometries, like the shear flow consider in this work. But if we pursue this path, the 3D equivalent to \eq{continuity} reads:
\begin{equation}
  \Sigma \bmath{v} = \nabla \Psi \bmath{\times} \nabla g,
\end{equation}
where $g(\bmath{x})$ is a second stream function.  In 2D $g=z$ recovers \eq{continuity}. In 3D, constants of $\Psi$ and $g$ constitute surfaces, and the intersections of two surfaces defines a streamline. Following the procedure similar to \se{streamf} we can obtain a set of diffusion equations, one for each component of the vorticity. Unfortunately, the diffusion equations are now functions of two scalar quantities ($\Psi$ and $g$) and therefore require a much more complex procedure to solve, probably involving an iterative approach. 

\subsection {Sensitivity to boundary conditions}
A quite general, and somewhat surprising, feature of the numerical model is its sensitivity to the outer boundary constraints (BC). Here, we have used the linear solution presented in \se{linear} to provide the flow velocity at the outer boundary. This turns out rather well, since the flow pattern is not very sensitive to the choice of $r_\mathrm{out}$, although we found that the atmosphere region converges only for large $r_\mathrm{out}$ (requiring high numerical resolution). However, if the flow quantities at the outer boundary is changed to the background flow pattern, $\Psi_\infty$, omitting $\psi_{y0}$--which means there is that $v_x=0$ by construction -- it is found that the flow pattern is very different. In some cases even retrograde rotation around the perturber is observed.

Since the linear solution that we presented in \fg{hs-orbits} agrees with previous works \citep[\eg][]{MassetEtal2006,PaardekooperPapaloizou2009}, we have confidence in the robustness of the nonlinear calculations that we derived in the later sections. However, we have neglected possible contributions of the spiral density wave, \ie\ from the Lindblad torque region. \citet{PaardekooperPapaloizou2009} considered the effect of the Lindblad torque (which originates at distances $x\gtrsim H$) on the flow in the vicinity of the protoplanet. They found that is was justified to neglect the Lindblad torque in the limit of zero softening of the gravitational potential (like in this work). Nevertheless, it is worthwhile to include the effects from the linear density wave theory \citep[\eg][]{Ward1997,TanakaEtal2002} as a boundary constraint, and to assess how this affects the flow structure near the planet.

\subsection{Feasibility of the laminar approximation}
The laminar approximation -- and the neglect of turbulence -- is another caveat. One stability criterion is the Richardson number, which measures the stability of a stratified flow:
\begin{equation}
  \mathrm{Ri} 
  = \frac{N^2}{\left( \partial v /\partial r \right)^2}
  = -\frac{g(r)}{\rho}\left( \frac{\partial \rho}{\partial r} \right) \left/ \left( \frac{\partial v_\theta }{\partial r} \right)^2 \right.
\end{equation}
where $N^2$ is the Brunt-V\"ais\"al\"a frequency. Richardson numbers below a critical value, $\mathrm{Ri}_\mathrm{crit} \simeq 1/4$, indicate that the kinetic energy term due to shear will outweigh the (stabilizing) term due to density stratification. The flow then overturns and becomes unstable. We have calculated the Richardson number for the simulations presented in \fg{mixed3}. In the far field ($r\gg 1$) it was found that $\mathrm{Ri}<\mathrm{Ri}_\mathrm{crit}$; but this is due to our initial setup, where $\Sigma_\infty$ is constant (allowing for gradients in $\Sigma_\infty$ and other background quantities [$\tilde{w}_\infty$, $c_\infty$] would be another obvious extension of the method.)  However, Ri increases with decreasing $r$ and closer to the planet $\mathrm{Ri}>\mathrm{Ri}_\mathrm{crit}$.

Other instabilities may still operate though. For example, the flow within the atmosphere may be convectively unstable. Flow past bodies characterized by large Reynolds numbers will generally become turbulent in the wake of the body. Vortices may develop along the horseshoe streamline \citep{KollerEtal2003}. Or the gas may just be simply unstable to start with, \eg\ due to the magneto-rotational instability \citep{BalbusHawley1991}. All these effects render the laminar approximation rather fragile. 

But it would be difficult, if not impossible, to present an analytic framework for non-steady flow. Our calculations provide the first order effect against which more specific simulations can be contrasted.  For example, \fg{p-traj} shows the first-order effect of the interaction of particle with gas. Turbulence  will change the particle trajectories; but, although their individual trajectories can be very different, it is yet to be shown that turbulent motions will result in a significant deviation to the \textit{time-averaged} accretion rate.

\subsection{Implications for protoplanet growth}
Notwithstanding the canonical scenario \citep{Safronov1969}, where protoplanetary cores are thought to be built up from big km-size (or larger) planetesimals, protoplanetary growth may in reality be driven by small particles, because:
\begin{enumerate}
  \item planetesimals will grind themselves down, before an embryo can accrete them \citep{KobayashiEtal2010,KobayashiEtal2011} or be trapped in resonances (and subsequently be ground down; \citealt{WeidenschillingDavis1985}).
  \item (sub)mm-observations imply that a large reservoir of the dust is typically observed in the $\sim$mm/cm-size range \citep[\eg][]{AndrewsWilliams2005}.
  \item growth by (big) planetesimals is slow. This is due to the negative feedback of a growing embryo on the planetesimal population (\ie\ it excites them to large eccentricity, rendering growth less efficient; \eg\ \citealt{KokuboIda1998,FortierEtal2007}).
\end{enumerate}
For these reasons, it is relevant to study in greater detail the interaction between small particle and embryo. In a follow-up work we will perform a quantitative (parameter) study to obtain the accretion potential of small particles.  However, from \fg{p-traj} two trends become clear. First, very small particles (\ie\ dust) do not make formidable building blocks. They just follow the streamlines of the gas, and will only accrete on to the planet if the gas does not so as well. Secondly, accretion rates rise steeply with particle size. In a recent study, \citet{LambrechtsJohansen2012} showed that a putative core accretes $\tau_p\sim0.1$ pebbles very rapidly with impact radii approaching the Hill radius, consistent with the findings in \fg{p-traj}c and f (the scale of the panels is the Hill radius). That this pebble accretion scenario results in high accretion rates is now widely recognized and may serve as an alternative to (the much slower) planetesimal accretion. However, one must not forget that $\tau_p=1$ particle also \textit{drift} the fastest \citep{Weidenschilling1977}; they may be removed from the system before they encounter a protoplanet \citep{KobayashiEtal2010}.  In this regard, somewhat smaller particles may offer a better channel for growth (see \citealt{OrmelKobayashi2012} for an assessment of the combined effects that affect the accretion behavior of small particles).

Protoplanets can also grow by contracting their atmospheres, allowing more gas to become bound. In this study this effect was crudely reflected by the transition density parameter $\Sigma_T$: the higher $\Sigma_T$ the further the isothermal regime penetrates, and the more massive the atmosphere.  More realistically, the atmosphere mass $M_A$ is a function of the (solid and gas) opacity, accretion rate, equation of state, \etc, and is obtained by solving the full equations of stellar structure \citep{MordasiniEtal2012}. These calculations are usually carried out in 1D, assuming radial symmetry.  But it is only through multidimensional (preferable 3D) studies that a true assessment of the \textit{boundary} between the atmosphere and nebula gas can be obtained \citep[\eg][]{LissauerEtal2009}. In this work we find that the headwind plays a critical role. A headwind renders the flow -- and therefore the atmosphere -- asymmetrical, see \eg\ \fg{mixed}c, which causes nebular material to penetrate the atmosphere within a fraction of the Bondi radius.  The consequences of the asymmetry and, in particular, the small atmosphere radius have yet to be addressed by protoplanet atmosphere models. We remark that, although the importance of the asymmetry increases with increasing $v_\mathrm{hw}$, it is a general feature -- seen also at low headwinds.


\section{Summary and conclusions}
\label{sec:conclusions}
In this manuscript the inviscid, steady state solution of a subsonic flow past gravitating bodies was considered.  First, in \se{linear}, an approximate but analytic solution was presented. This solution only holds in the linear regime, that is, at distances much larger than the Bondi radius $R_b$. The linear solution served as a boundary condition for the full (non-linear) numerical calculations (\se{simul-res}).  These showed the importance of the atmosphere region (a nonlinear phenomenon), where the flow curls around the planet in the prograde direction. Using the findings from the numerical simulations, we next constructed a more complete analytical approximation of the flow pattern, which captures the dynamics of the atmosphere region (\se{approx}). This solution very well matches the numerical findings and may be used in subsequent studies for which the flow pattern at scales of $R_b$ is important.

The adopted framework is quite general, with parameters describing the mass of the perturber, the headwind velocity $v_\mathrm{hw}$, and the shear parameter $\omega_\infty$.  Apart from \se{hw-only}, this study has focused on the flow pattern past small planets embedded in Keplerian discs ($w_\infty =-3\Omega_0/2$). However, flow patterns around `bodies' occur in diverse astrophysical settings, \eg\ in (small) satellites embedded circumplanetary disks \citep[\eg][]{EstradaEtal2009}, star formation \citep{KrumholzEtal2005}, or accretion disks around black holes \citep{mcKernanEtal2011}. As long as these embedded objects do not open gaps and the system is in steady-state, the framework constructed in this paper also provides a description for the flow pattern in these environments.


The key findings of this study are the following:
\begin{enumerate}
  \item For flows without shear, the steady-state flow pattern resembles that of an hourglass (\fg{headw}), where gravity focuses the streamlines.
  \item For Keplerian shear flows the vorticity, which is amplified near the planet due to conservation of vortensity, plays the dominant role. There is a point in the simulation where the flow stagnates. In general there are two solutions for the  associated stagnation streamline -- one defining the atmosphere region and one defining the horseshoe region -- which only coincide when $v_\mathrm{hw}=0$.  
  \item For the (general) $v_\mathrm{hw}\neq0$ case, the boundary of the atmosphere is asymmetric and lies within the Bondi sphere (\fg{atm}). The size of the atmosphere reduces with increasing headwind but its averaged radius is best approximated by the Bondi radius $R_b$ (rather than the Hill radius).
  \item When the flow is subsonic (in particularly, $v_\mathrm{hw}\ll 1$), the density near the Bondi radius can be well approximated by the hydrostatic solution, such that $\Sigma$ becomes a function of radius only. In this approximation, an analytic solution for the stream function has been obtained (see \se{approx} and \app{full-sol}). This solution can be used to compute the gas drag force which solid particles experience during their encounter with the planet.
  \item When they can penetrate the atmosphere region, particles can be accreted as they settle to the planet through the circumplanetary disc. Accretion of small particles (dust) is suppressed as they strongly couple to the gas flow, but we find that for $\sim$mm-size and larger particles accretion rates should be high. There is also tentative evidence that a (moderate) headwind accelerates particle sweepup, due to the smaller atmosphere.
\end{enumerate}

\section*{Acknowledgments}
This work has profited immensely from discussion with many colleagues,
including: Eugene Chiang, Gennaro, D'Angelo, Kees Dullemond, Anders Johansen, Hiroshi Kobayshi, Don Korycansky, Christoph Mordasini, Ruth Murray-Clay, Satoshi Okuzumi, Ryan O'Leary, Sijme-Jan Paardekooper, Jiming Shi, Hidekazu Tanaka, Takayuki Tanigawa, Neal Turner, and others. Special thanks goes to the referee, Hagai Perets, for providing a very helpful review. Support for this work was provided by NASA through Hubble Fellowship grant \#HST-HF-51294.01-A awarded by the Space Telescope Science Institute, which is operated by the Association of Universities for Research in Astronomy, Inc., for NASA, under contract NAS 5-26555.


\appendix
\section[]{Derivation of Equations (29), (30)}
\label{app:linear}
The linearization of \eq{D-nondim} proceeds as follows. The term $\nabla\Psi/\Sigma$ is replaced by $\nabla (\Psi_\infty +\psi) /(1+\sigma) \approx \nabla\psi +(1-\sigma)\nabla\Psi_\infty$, where it is assumed that $\sigma\ll1$ and that we can neglect terms like $\sigma \nabla\psi$. Since $\nabla\Psi_\infty = (-\omega_\infty x +{\cal M}_\mathrm{hw})\mathbf{e}_x$ only has an $x$-component, the divergence operator $\nabla \cdot$\ is replaced by a differential $\partial_x$. In this way \eq{D-linear} becomes
\begin{equation}
  \nabla^2 \psi -(\partial_x\sigma)(\omega_\infty x -{\cal M}_\mathrm{hw}) 
  -\omega_\infty(1-\sigma)
  = -\omega_\infty -\tilde\omega\sigma.
\end{equation}
from which, after re-arranging, \eq{D-linear} is retrieved.

For the linearization of \eq{B-nondim}, an isothermal EOS is assumed, such that $W(\Sigma) = \log \Sigma \approx \sigma$. The kinetic term $|\nabla\Psi|^2/2\Sigma$ is linearized as
\begin{align}
 \nonumber
  \frac{(\nabla\Psi_\infty +\nabla \psi)^2}{2(1+\sigma)^2}
  \approx
  \frac{1}{2}|\nabla\Psi_\infty|^2 -\sigma|\nabla\Psi_\infty|^2 +\nabla\Psi_\infty \cdot\nabla\psi \\
  =
  \frac{1}{2}|\nabla\Psi_\infty|^2 +({\cal M}_\mathrm{hw} -\omega_\infty x)(\partial_x\psi) -\sigma({\cal M}_\mathrm{hw} -\omega_\infty x)^2
\end{align}
On the RHS of \eq{B-nondim} $\Psi = \Psi_\infty +\psi$. After noting that the first order terms cancel (they obey the far-field solution), \eq{B-linear} is obtained.

\section[]{Full, nonlinear expressions for the flow pattern}
\label{app:full-sol}
In \se{approx} an approximate solution for the flow in terms of the stream function (see \eqp{continuity}) was presented in terms of an homogeneous solution, $\Psi_\mathrm{hom}$, and a particular solution, $\Psi(r)$: $\Psi(r,\theta) = \Psi_\mathrm{hom}(r,\theta) +\Psi(r)$. We found that (\eqp{Dpsi-r}):
\begin{equation}
  \label{eq:Dpsir}
  \Psi'(r) =
  \frac{e^{\frac{m}{r}} \left[ 
      m^2\tilde\omega\text{Ei}\left(\frac{m}{r}\right)
      - \tilde\omega r(m+r)  e^{\frac{m}{r}}
      + 2(r^2 +C_1) 
      \right]}{2 r},
\end{equation}
where $C_1$ is an integration constant, and $\text{Ei}(x)$ the exponential integral
\begin{equation}
  \mathrm{Ei}(x) = \int_{-\infty}^x dt\ \frac{e^t}{t}.
  \label{eq:expEi}
\end{equation}
\Eq{Dpsir} can be integrated once more
\begin{align}
  \nonumber
  \Psi_r(r) = 
  \frac{1}{4} \left[-\text{Ei}\left(\frac{m}{r}\right) \left(4 C_1+m^2
   \tilde\omega \text{Ei}\left(\frac{m}{r}\right)+2 m^2\right)
   \right.
   \\ \nonumber
   +4C_2+8 m^2 \tilde\omega \text{Ei}\left(\frac{2 m}{r}\right)
   \\
   \left.
   +r e^{m/r} \left(2
   (m+r)-\tilde\omega e^{m/r} (4 m+r)\right)\right]
  \label{eq:psir-full}
\end{align}

These expressions contain two integration constant, $C_1$ and $C_2$. As argued in \se{approx}, these are chosen such that they match the linear solution at a point $(x,y)=(0,y_c)$ or $r=y_c$ and $\theta=\pi/2$ in radial coordinates. This resulted in conditions given by \eqs{psi-match}{Dpsi-match}. The derivative of \eq{psi0}, $\psi_{y0}$ reads
\begin{equation}
  \partial_y \psi_{y0} = m \left[ \cos \tilde{y} \Ci(\tilde{y}) -\frac{1}{2}\sin \tilde{y} \left( \pi -2\Si(\tilde{y}) \right) \right],
  \label{eq:Dpsiy0}
\end{equation}
where $\tilde{y} = y/\sqrt{2}$, and $y>0$ and a Keplerian disk ($\tilde\omega=1/2$) have been assumed.

The constants $C_1,C_2$ follow by rearranging \eqs{psi-match}{Dpsi-match}, \ie\ by isolating $C_1$ and $C_2$.  This is a straightforward procedure, but rather cumbersome.  For a Keplerian disk ($\tilde\omega=1/2$) we obtain in this way
\begin{align}
  \nonumber
  C_1 = 
  e^{-\frac{m}{y_c}} &\left\{ m y_c \Ci\left(\tilde{y_c}\right) \cos
   \tilde{y_c} -e^{\frac{m}{y_c}} \left[\frac{1}{4} m^2
   \Ei\left(\frac{m}{y_c}\right) +y_c^2\right] \right. \\ \nonumber
&  -m y_c \left[\frac{\pi}{2} - \Si\left(\tilde{y_c}\right)\right]\sin\tilde{y_c}  \\
&  \left.
   +\frac{1}{4} y_c e^{\frac{2 m}{y_c}}
   \left(y_c+m\right)+\frac{1}{4} y_c \left(3 y_c+m\right)\right\}.
  \label{eq:C1-full}
\end{align}
An expression for $C_2$ can be obtained in a similar way.  In the limit of $m\ll y_c\ll1$ these expressions simplify significantly and read:
\begin{align}
  C_1 \approx &\ m y_c \left(\log \left(\frac{y_c}{\sqrt{2}}\right) +\gamma \right) \\
  C_2 \approx &\ \frac{1}{24} m \left\{12 \sqrt{2} \pi +4 y_c 
  \left[\log \left(\frac{y_c^6}{8}\right) +6 \gamma -6 \right. \right.
  \\ 
 &\ \left. \left. +\left(\log \left(\frac{y_c^6}{8}\right)+6 \gamma \right) 
   \left(-\log \left(y_c\right)+\log (m)+\gamma \right)
   \right] \right\} \
   \nonumber
\end{align}
where $\gamma\approx0.577$ is the Euler-Mascheroni constant.

To conclude, let us summarize the expressions for the flow velocity that follow from the full nonlinear model. For the radial component it reads:
\begin{equation}
  \label{eq:flow-field-r}
  v_r
  = \frac{1}{\Sigma} \frac{\partial_\theta \Psi}{r}
  = \frac{1}{\Sigma r} \partial_\theta \Psi_\mathrm{hom},
\end{equation}
where $\Sigma = \exp(m/r)$, and $\Psi_\mathrm{hom}$ is given by \eq{psihom} which includes the $c_1, c_2$ integration constants. The azimuthal flow velocity reads 
\begin{equation}
  \label{eq:flow-field-phi}
  v_\theta
   =-\frac{1}{\Sigma} \partial_r \Psi
   =-\frac{1}{\Sigma} \left[ \partial_r \Psi_\mathrm{hom} +\Psi'(r) \right]
\end{equation}
and includes in addition the integration constant $C_1$ through the $\Psi'_r$ term (\eqp{Dpsir}).

\bsp
\label{lastpage}
\end{document}